\documentclass{aa}

\bibpunct{(}{)}{;}{a}{}{,}

\usepackage{booktabs}
\usepackage{threeparttable}
\usepackage[table]{xcolor}
\usepackage{graphicx}
\usepackage{color}
\usepackage[normalem]{ulem}
\usepackage{txfonts}
\newcommand{\ra}[1]{\renewcommand{\arraystretch}{#1}}

\newcommand\redsout{\bgroup\markoverwith{\textcolor{red}{\rule[0.5ex]{2pt}{0.4pt}}}\ULon}

\begin{document}

   \title{Matching LOFAR sources across radio bands}

   \author{L. B\"ohme\inst{1}\thanks{E-mail: \href{mailto:Lboehme@physik.uni-bielefeld.de}{Lboehme@physik.uni-bielefeld.de}}
          \and
          D. J. Schwarz\inst{1}
          \and 
          F. de Gasperin\inst{2,3}
          \and
          H. J. A. R\"ottgering\inst{4}
          \and
          W. L. Williams\inst{4,5}
          }

   \institute{Fakultät für Physik, Universität Bielefeld,
              Postfach 100131, 33501 Bielefeld, Germany
         \and 
             Istituto di Radioastronomia, Istituto Nazionale di Astrofisica,
             Via Gobetti 101, 40127 Bologna, Italy         
        \and
             Hamburger Sternwarte, Universität Hamburg,
             Gojenbergsweg 112, 21029 Hamburg, Germany
        \and
            Leiden Observatory, Leiden University, 
            PO Box 9513, NL-2300 RA Leiden, the Netherlands 
        \and 
            SKA Observatory, Jodrell Bank, Lower Withington, Macclesfield, SK11 9FT, United Kingdom
            }

   \date{Received December 12, 2022; accepted April 27, 2023}

 
  \abstract
   {}
      {
    With the recent preliminary release of the LOFAR LBA Sky Survey (LoLSS), the first wide-area, ultra-low frequency observations from LOFAR were published. Our aim is to combine this data set with other surveys at higher frequencies to study the spectral properties of a large sample of radio sources.}
   {We present a new cross-matching algorithm taking into account the sizes of the radio sources and apply it to the LoLSS-PR, LoTSS-DR1, LoTSS-DR2 (all LOFAR), TGSS-ADR1 (GMRT), WENSS (WSRT) and NVSS (VLA) catalogues\thanks{Full catalogue as described in Table \ref{tab:catalogue} and the sub-catalogues described in Sect.~\ref{subsec:cat} are available at the CDS via anonymous ftp to \url{cdsarc.u-strasbg.fr} (\url{130.79.128.5}) or via \url{http://cdsarc.u-strasbg.fr/viz-bin/cat/J/A+A/}}. We then study the number of matched counterparts for LoLSS radio sources and their spectral properties.}
   {We find counterparts for 22\,607 (89.5\%) LoLSS sources. The remaining 2\,640 sources (10.5\%) are identified either as an artefact in the LoLSS survey (3.6\%) or flagged due to their closeness to bright sources (6.9\%). We find an average spectral index of $\alpha = -0.77 \pm 0.18$ between LoLSS and NVSS. Between LoLSS and LoTSS-DR2 we find $\alpha = -0.71 \pm 0.31$. The average spectral index is flux density independent above $S_{54} = 181$ mJy. Comparison of the spectral slopes from LoLSS--LoTSS-DR2 with LoTSS-DR2--NVSS indicates that the probed population of radio sources exhibits evidence for a negative spectral curvature.}
   {}


   \maketitle
%

\section{Introduction}\label{Sec1}


Radio continuum surveys map areas of the 
sky at various depths, resolution and frequency. The 
cross-identification of sources from surveys at different radio frequencies and other wavelengths 
is a key element in the astrophysical interpretation of 
radio sources. 

For the simplest cross-matching task, a single emission mechanism dominates, e.g.~synchrotron (radio) or thermal dust emission (infrared). In that case the general properties of the spectral energy distribution (SED) are known, i.e.\ power-law or black-body spectrum, and the task is to measure the flux density at different frequencies and to determine a small number of parameters that describe the SED (e.g.\ spectral index or physical temperature). In contrast, cross-matching of sources involving multiple emission mechanisms is more complicated. 
It requires the full multi-wavelength (and sometimes multi-messenger) astronomy. This is for example the case when matching radio with optical and infrared data. Here, the general form of the SED is unknown and the task is to identify sources across different wavelengths.

In this work, we are concerned with the synchrotron regime, as we cross-match continuum radio surveys from the lowest radio frequencies as observed with the Low Frequency Array 
\citep[LOFAR;][]{LOFAR}, up to $1.4$ GHz. 
We use a new algorithm for radio-radio cross-matching of catalogues, which also incorporates the source extension of resolved sources, and apply it to radio catalogues from LOFAR, the Giant Metrewave Radio Telescope (GMRT), the Westerbork Synthesis Radio Telescope (WSRT), and the Very Large Array (VLA). 
The corresponding radio surveys are the LOFAR LBA Sky Survey \citep[LoLSS;][]{deGasperin2021}, the LOFAR Two-meter Sky Survey \citep[LoTSS;][]{LoTSS1,LoTSS2}, the TIFR GMRT Sky Survey alternative data release 1\footnote{We will use TGSS as an abbreviation for TGSS ADR1} \citep[TGSS ADR1;][]{TGSS},
the Westerbork Northern Sky Survey \citep[WENSS;][]{WENSS}, and the NRAO VLA Sky Survey \citep[NVSS;][]{NVSS}.
These radio surveys were chosen, as they all cover the Hobby-Eberly Telescope Dark Energy Experiment \citep[HETDEX;][]{Hill2008} Spring Field and have a broad range in frequencies from 54 to 1\,400 MHz. The angular resolutions of the surveys are between $6''$ and $54''$. The final cross-matching catalogue includes 22\,624 LoLSS sources and is publicly available. 

Except for the deepest observations, most of the observed objects in the radio sky are active galactic nuclei (AGNs) for which synchrotron radiation from relativistic electrons (CREs) is the origin of the radio emission \citep{best2021,degasperin2018}. To gain insight into the physical phenomena that produce the emission, the characteristics of the SED in the radio band is an important tool. Generally, the radio SEDs are smooth over a large frequency range and often a single power law can be used to describe the SEDs (see e.g.\ \citet{Kellermann1969}).

Radio-selected AGNs sample a small fraction (<10\%) of all AGNs, with a bias towards bright and jetted AGNs \citep{padovani2017}. In jetted AGNs there are four types of regions, a core, jet(s), hot spot(s) and lobe(s). Often a binary classification based on the morphology of emission is possible \citep{FR}. Herein Fanaroff-Riley class I (FR I) describes centre-brightened sources with jets disrupting on the kpc scales to subsonic plumes. For edge-brightened AGNs (FR II) the jets remain relativistic throughout, escaping the galactic nucleus with little deceleration and eventually each ending in a hot spot \citep{mingo2019}. FR II sources are generally brighter, though a low-luminosity population (FR II-Low) was recently found, with three orders of magnitude lower luminosities \citep{mingo2019}. Both classes can also be used to describe the different jet interactions with the environment \citep{blandford2019}. Relativistic jets (FR II) create hot cocoons that protect them from destructive instability, while FR I sources directly heat their galactic surroundings. These FR I plumes can still propagate several Mpc away from their host and the heating can stimulate star formation and may therefore play a major role in the evolution of galaxies \citep{blandford2019}. 
For both classes a flat spectrum core is found which may result from synchrotron self-absorption. 

The \textit{spectral index} $\alpha$ is defined as the slope of the radio SED in a log-log plot,
\begin{equation}
\alpha (\nu) = \frac{\mathrm{d} \log S_\nu}{\mathrm{d} \log \nu},
\end{equation}
with the flux density $S_\nu$.
The spatial variation of the spectral index across extended sources is fundamental in understanding the different mechanisms of radio emission in galaxies \citep{degasperin2018}. SED studies of radio galaxy lobes, which generally show a negative curvature, with a steeper spectrum at high frequencies and with older age, can give insight into the source age, energetics and plasma composition \citep{degasperin2018}.
At low frequencies, the lobes are the dominant flux contributing component.
When integrating the flux into a single point (as for unresolved sources) the lobes therefore play an important role in the measured curvature of the spectral index. Thus even the measurement of curvature in unresolved sources should give insight into the morphology of these.
To estimate the spectral index or SED of a radio source, measurements at different frequencies are needed. Since intensity, resolution, and also astrometric precision cannot be expected to agree at different frequencies, cross-matching of radio sources at different frequencies must ensure that components (or sources) are matched properly.

A fast and easy approach to measure spectral indices in the sychrotron emission regime from source catalogues at different radio frequencies is the positional cross-matching \citep{degasperin2018}. With a fixed search radius around each source in a catalogue, counterparts in other catalogues at different frequencies are identified. A caveat of this approach is that resolution differs from survey to survey. Multiple sources in a high angular resolution survey can appear blended into a single source at lower resolution. For surveys at different frequencies, the sources might also appear different, as some source components may not be visible at the lower or higher frequency.

Several improved algorithms were presented in the literature. In the following a short and incomplete summary is given:
The general use of a Likelihood Ratio (LR) was discussed by \citet{sutherland1992}, summarising previous works and deriving the reliability of radio with optical associations.
A general approach was developed by \citet{budavari2008} for cross-identifying point sources in multiple observations based on a Likelihood Ratio (LR) using a Bayesian approach.
They assumed astrometric precision to be the limiting factor. This method gives a more refined cross-matching than the simple positional search based on the assumption that sources have a single component and are point-like (or unresolved), however it is not suitable for extended and asymmetric sources. 
An updated version of this LR matching was presented by \citet{fan2020}, which also takes into account the lobe morphology and different angles between them. This solves several drawbacks of the previous model, but is still only used for cross-matching extended radio sources with optical/IR point-sources. It can be used after the approach of this paper to cross-identify the optical/IR counterparts for the highest-resolution radio survey and then propagate these results to the other matched radio surveys.
A more specific approach is the positional update and matching algorithm \citep[PUMA;][]{line2017} which was developed for cross-matching low-frequency radio catalogues and combines the general approach by \citet{budavari2008} with a spectral matching criterion and the possibility for multiple counterparts.
It first does a positional match between a base catalogue and all others, then isolated matches (having a 1:1 assignment) are accepted if either their assigned matching probability by the algorithm is above some chosen threshold or if a power-law spectral model fits. For non-isolated matches (so for several possible candidate matches in one survey for one source in the base survey), if one has a much better power-law fit, it is accepted. If this is not the case, a combination of all single sources with combined flux density is tested for a power-law fit. Overall, the PUMA algorithm gives a better positional match, especially when the source density is high and multiple possible counterparts are nearby. 
The advantage of LR algorithms is the handling of multiple potential matches \citep{sutherland1992}. In the method presented in this work, every match above some threshold ends up in the catalogue, while LR algorithms account for the possibility that the match sits at a completely different redshift. As only radio surveys are considered in this work and LoLSS sources are dominantly AGNs, this is not an important issue, as the source density is still low enough. However, for deeper fields like the LOFAR Deep Fields \citep{best2021}, future radio surveys and matching with optical surveys, this will be an issue.

The new algorithm that is presented here presents an improvement over a simple positional cross-matching in a fixed search radius and attempts to take position, shape and orientation of radio sources into account. It currently provides a binary answer (match or no match), which may be generalised to produce a posterior probability distribution.

In Sect.~\ref{Sec2} of this work, we introduce the radio surveys used for this study. Section \ref{Sec3} describes the cross-matching algorithm, how the resulting catalogue was cleaned, and its properties. Section \ref{Sec4} is about the dependency between the spectral index and flux density, redshift, and source compactness. In Sect.~\ref{Sec5} we end with suggested future updates to the algorithm and conclude on our finding.

\section{Data}\label{Sec2}

The spectral index catalogue is created from the following five radio surveys:

\textbf{LoLSS} -- Preliminary Release \citep[PR;][]{deGasperin2021}: It covers 740 deg$^2$ in the northern sky around the HETDEX Spring Field. The observations for the survey were taken with the low band antennas (LBA) using the Dutch stations of LOFAR in the frequency range $42$ -- $66$ MHz. For these observations the `LBA OUTER' mode was used, whose primary beam full width at half maximum (FWHM) is similar to the HBA counterpart at 144 MHz. The resolution of LoLSS PR is $47''$ due to the restrictions of direction-independent calibration, 
with a root mean square (rms) noise of about 5 mJy beam$^{-1}$ and an astrometric accuracy of $2.5''$.

\textbf{LoTSS} -- Data Release 1 and 2 \citep[DR1 / DR2;][]{LoTSS1,LoTSS2}: The first data release covers a 424 deg$^2$ region in the northern sky around the HETDEX Spring Field. The second data release covers a significantly larger portion of the northern sky --- $5\,635$ deg$^2$. Both surveys were taken with the high band antennas (HBA) using the Dutch stations of LOFAR. Both releases have a resolution of $6''$, with a median sensitivity of 71 and 83 $\mu$Jy beam$^{-1}$ for DR1 and DR2, respectively. The astrometric accuracy for both releases is $0.2''$. For the first data release, the LOFAR Radio Galaxy Zoo results were used to combine multiple components as jets, lobes, etc. present as separated entries in the source catalogue into single entries in the \textit{value-added catalogue} \citep{williams2019}. Also artefacts were removed and, if available, redshift estimate (most of them photometric), infrared and optical identifications \citep{duncan2019,duncan2021} from Pan-STARRS and WISE were added.

\textbf{NVSS} \citep{NVSS}: Its sky coverage includes the whole sky north of declination $-40^\circ$, therefore covering 82\% of the whole sky at a central frequency of 1\,400 MHz. It was observed with the VLA in New Mexico in D and DnC configurations. The resolution is $45''$ with an average rms noise of 0.45 mJy beam$^{-1}$. The VLA configuration and snapshot duration changed with declination, though this happens below the sky coverage used in this study and does not affect our results.

\textbf{TGSS} -- Alternative Data Release 1 \citep[ADR1;][]{TGSS}: It covers the full sky north of Dec $-53^\circ$ (90\% of the whole sky) at the central frequency of 147 MHz. It was observed with the Giant Meterwave Radio Telescope (GMRT) in India with a final resolution of $25''$ and a rms noise of 3.5 mJy beam$^{-1}$.

\textbf{WENSS} \citep{WENSS}: It covers the full sky north of 30$^\circ$ declination at a frequency of 325 MHz. It was observed with the Westerbork Synthesis Radio Telescope (WSRT) in the Netherlands. The resolution is declination dependent and is given as $54'' \times 54''/\cos(\delta)$. The rms noise is 3.9 mJy beam$^{-1}$.

\begin{figure}
    \resizebox{\hsize}{!}{\includegraphics{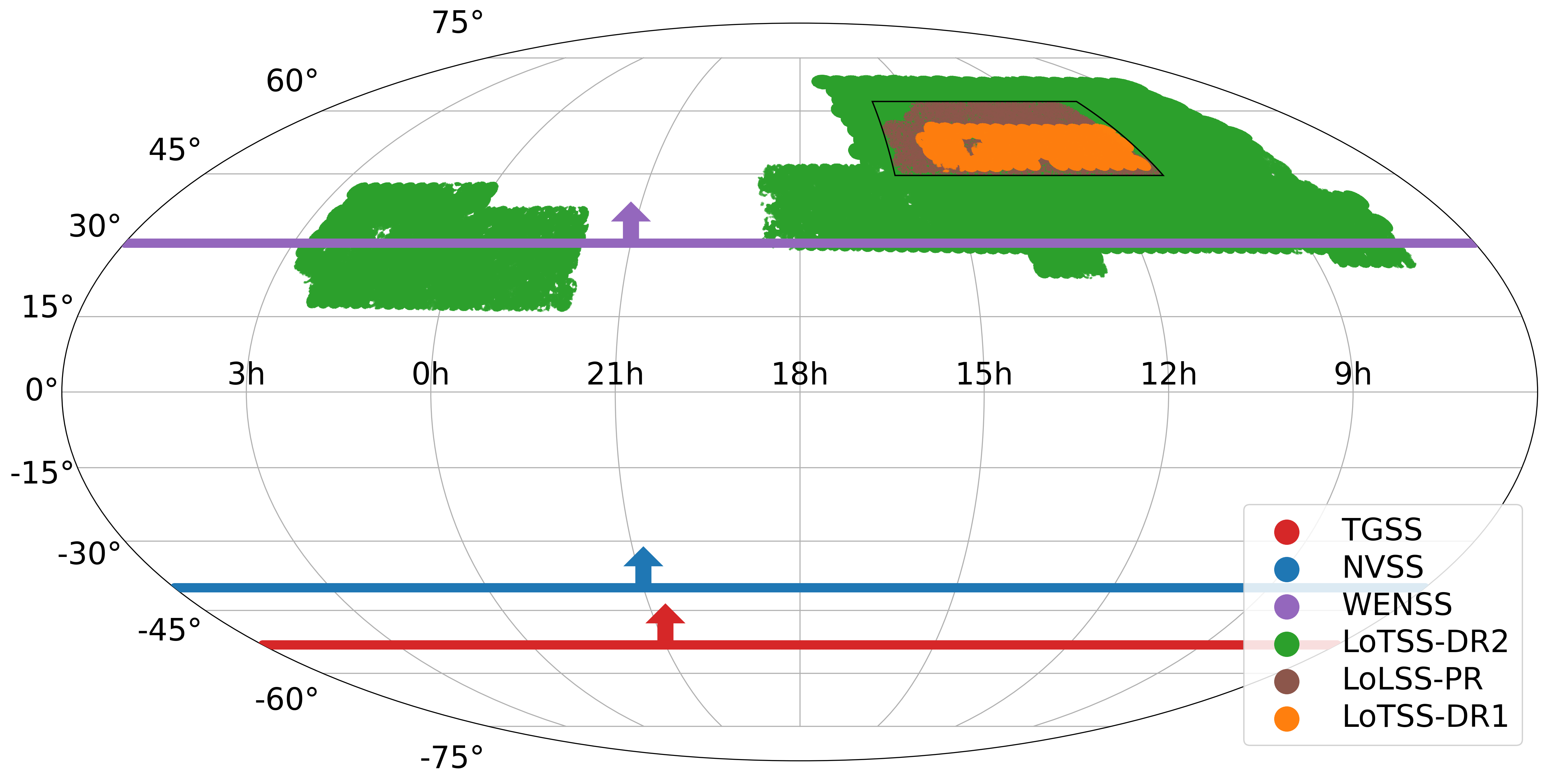}}
    \caption{Coverage map of the radio surveys used in this work. A line indicates the (southern) survey coverage boundary, where the area of observation is in the direction of the arrow pointing north. The black quadrangle around the survey footprint of LoLSS-PR marks the cross-matching region of this study, which is the HETDEX Spring Field.}
    \label{fig:overview_map}
\end{figure}

An overview of the most important characteristics of the surveys is also given in Table \ref{tab:overview} and their sky coverage is shown in Fig.~\ref{fig:overview_map}.

\begin{table*} \ra{1.3}
\centering 
\caption{Overview of the used surveys.}
\label{tab:overview}
\rowcolors{2}{lightgray!40}{}
\begin{tabular}{c r r r c c c c} \toprule
Survey name & Frequency & Resolution & rms noise & Sky area & Source density & Release & Reference\\
            &   [MHz]   & [arcsec] & [ mJy $\mathrm{beam}^{-1}$] & [$\mathrm{deg}^2$] & [$\mathrm{deg}^{-2}$] & &\\ \midrule
LoLSS-PR & 54 & 47 & 5 & 740 & 34.1 & 2021 & \citet{deGasperin2021} \\
LoTSS-DR1 & 144 & 6 & 0.071 & 424 & 765.6 & 2019 & \citet{LoTSS1} \\
LoTSS-DR2 & 144 & 6 & 0.083 & 5\,635 & 780.2 & 2022 & \citet{LoTSS2} \\
TGSS-ADR1 & 147 & 25  & 3.5 & 36\,900 & 16.9 & 2016 & \citet{TGSS}\\
WENSS & 325 & $54\times54/\cos(\delta)$ & 3.9 & 10\,177 & 22.5 & 2000 & \citet{WENSS} \\
NVSS & 1\,400 & 45 & 0.45 & 33\,000 & 53.7 & 1998 & \citet{NVSS}\\ \bottomrule
\end{tabular} 
\end{table*}

\section{Cross-matching}\label{Sec3}

\subsection{Cross-matching method}

The most important consideration when cross-matching these six catalogues is their different resolutions of $6''$ (for two surveys), $25''$, $45''$, $47''$ and $54''$. Sources that may be resolved into multiple sources or multi-component sources at a high resolution, may appear as a single resolved or even a single unresolved source at lower resolution. Therefore the source size is an important parameter to consider. In order to include this parameter, a geometrically based search algorithm was developed and used. It takes one source in the base catalogue – in this work LoLSS, as it has the lowest frequency, a constant low angular resolution and should contain many lobes – and compares it with all sources within the chosen search radius in the other catalogues. As a result, these are either accepted as matches or discarded. The multi-frequency-catalogue is then formed from the accepted matchings\footnote{An implementation of the algorithm can be found at \url{https://github.com/lboehme/cross-matching-lofar}}.

\begin{figure}
    \resizebox{\hsize}{!}{\includegraphics{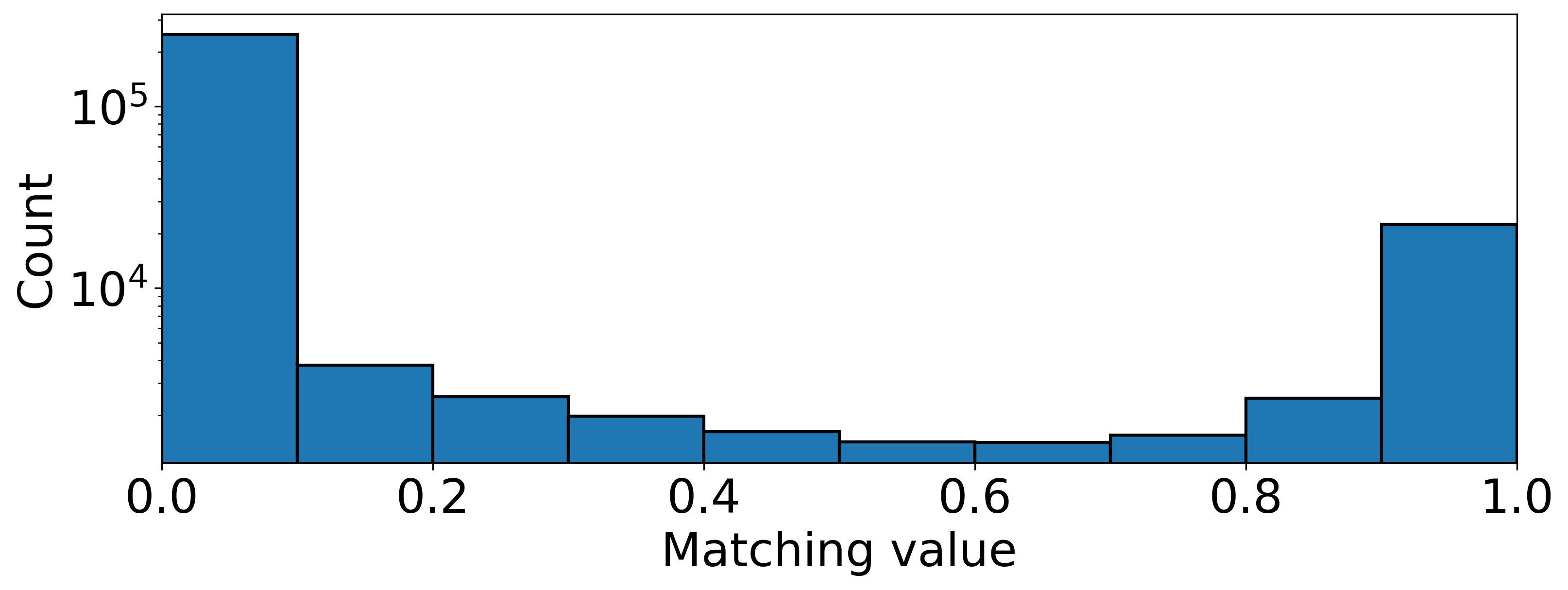}}
    \caption{Distribution of the Gaussian matching values $v$ for all potential matches calculated at their centre.}
    \label{fig:v_values}
\end{figure}

\begin{figure}
    \resizebox{0.8\hsize}{!}{\includegraphics{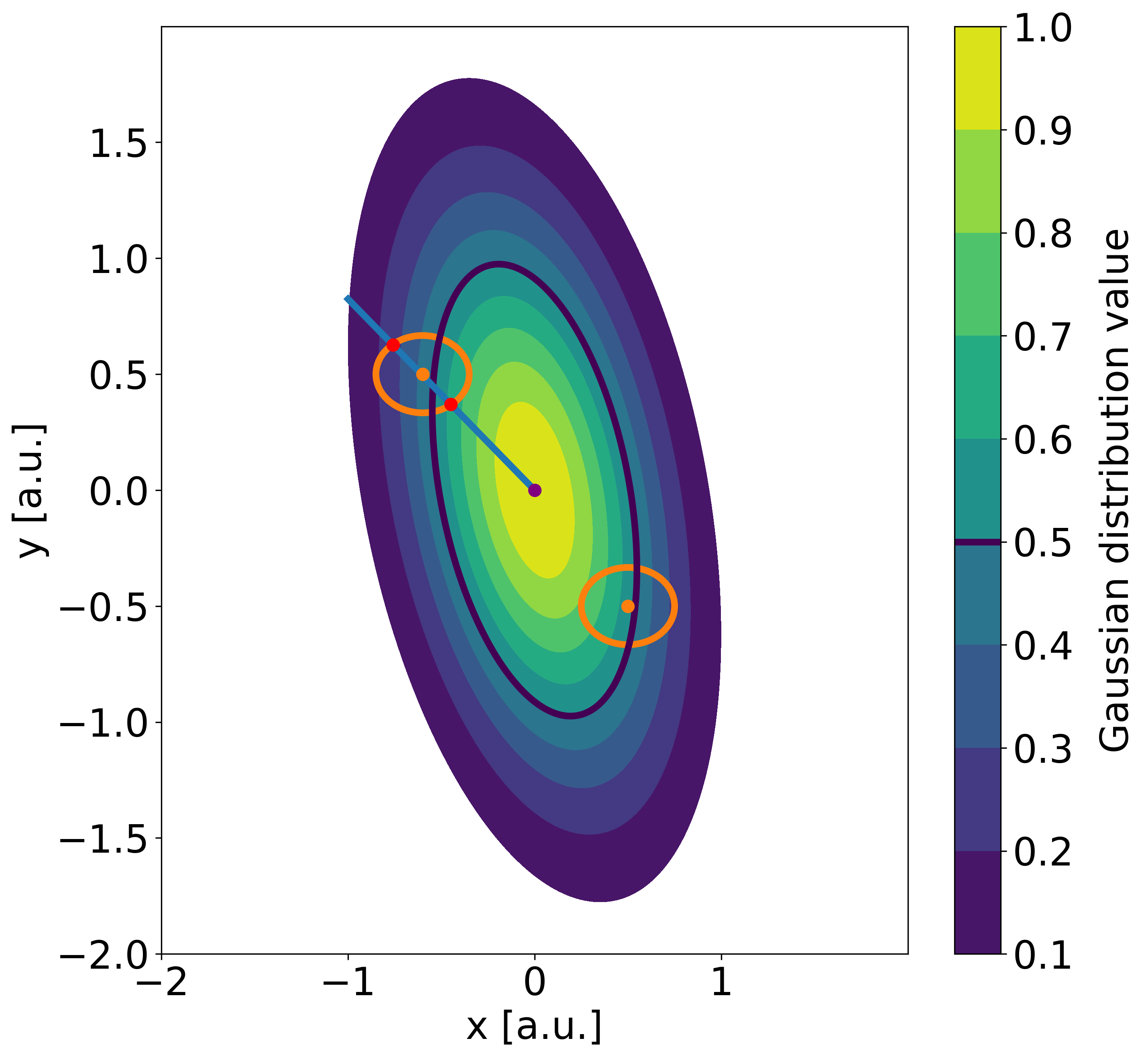}}\centering
    \caption{Sketch of three sources matching. The original (LoLSS) source in the middle is displayed as filled contours with its Gaussian distribution in 0.1 steps from 0 (Minimum) to 1 (Maximum). The potential counterpart sources are shown with their FWHM in orange. The blue line connects both centres and the red points are where the Gaussian value $v$ is additionally calculated.}
    \label{fig:matching}
    \resizebox{0.8\hsize}{!}{\includegraphics{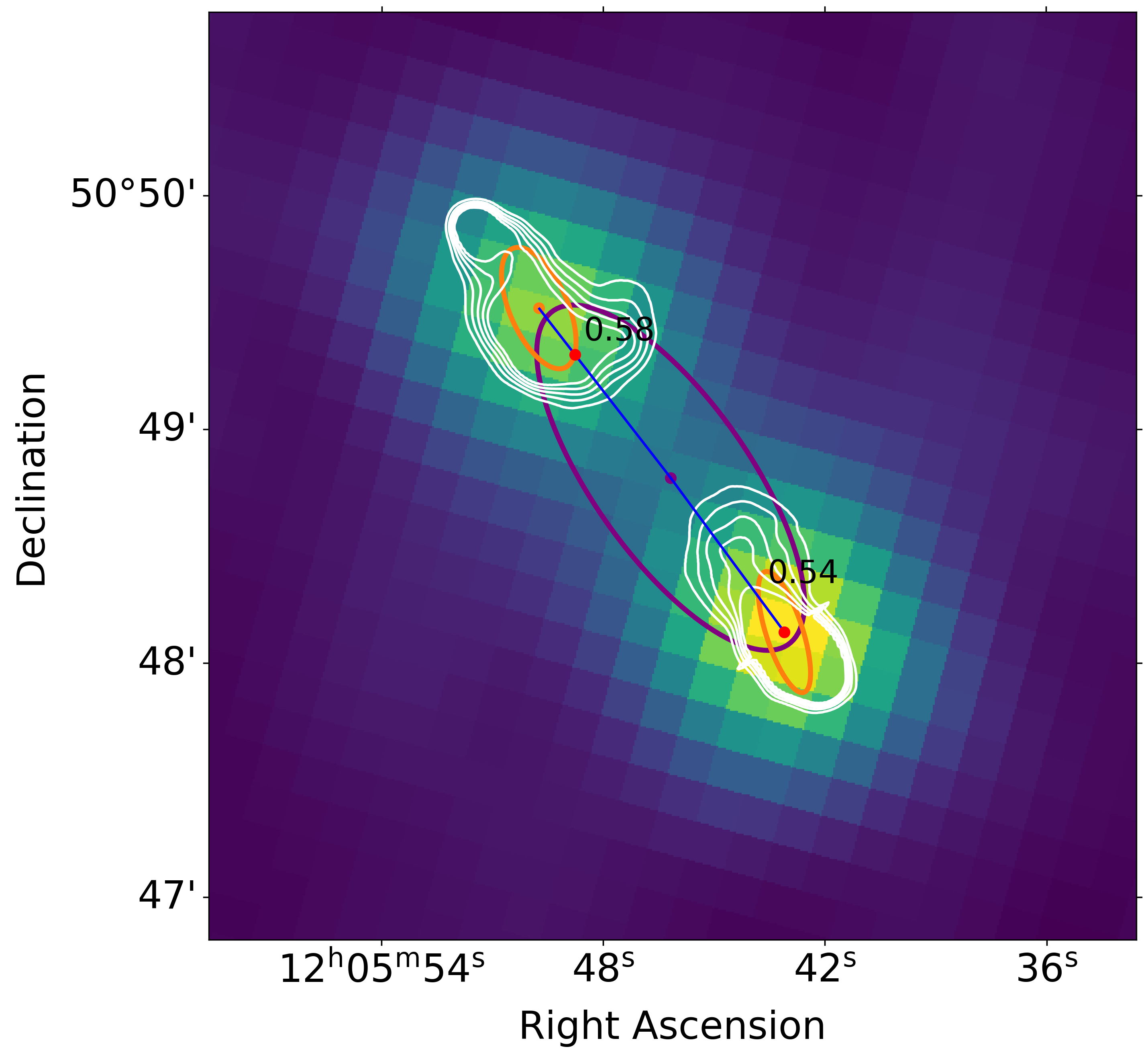}}\centering
    \caption{Example of two sources in LoTSS-DR2 (orange and white contours) matching with a LoLSS (purple outline and colour background) source. The orange and purple ellipses represent the FWHM, the blue line connects the centres, while the red dot is the point, where the value exceeds the matching condition and is accepted. Both matches are accepted, because either the second matching condition (upper left source) or the first (lower right source) is fulfilled, as indicated by the values.}
    \label{fig:matching_b}
\end{figure}

As a preparation, 2-D Gaussian functions $G(x,y)$ are created from the Major Axis (MA), Minor Axis (MI) and Positional Angle (PA) for all sources from the different catalogues. These use the fitted Gaussian distribution from the source extraction in PyBDSF \citep{pybdsf} and describe the FWHM ellipse at the $0.5$ (half maximum) level. As a first step, the peak-normalised 2-D Gaussian (maximum equal to 1) of a LoLSS-source is taken and its value $v = G(x_{cp},y_{cp})$ calculated at the centres of the potential counterparts. If this value is above some threshold value $v_{thr}$ in the interval $(0,1)$, the pair is accepted. For this analysis, $v_{thr}$ is chosen as $0.5$, as that describes the FWHM ellipse. When looking at the distribution of the centre matching values between all potential matches, shown in Fig.~\ref{fig:v_values} for the LoLSS and LoTSS-DR2 catalogues, most values are either close to 0 or 1. Therefore the impact of the precise choice of the threshold at values between 0.5 and 0.7 is small. 
An example for an immediate match is given in Fig.~\ref{fig:matching} and shown as the lower right orange contour. Here, the centre is inside the FWHM ellipse and therefore $v > v_{thr}$.

\begin{figure}
    \resizebox{\hsize}{!}{\includegraphics{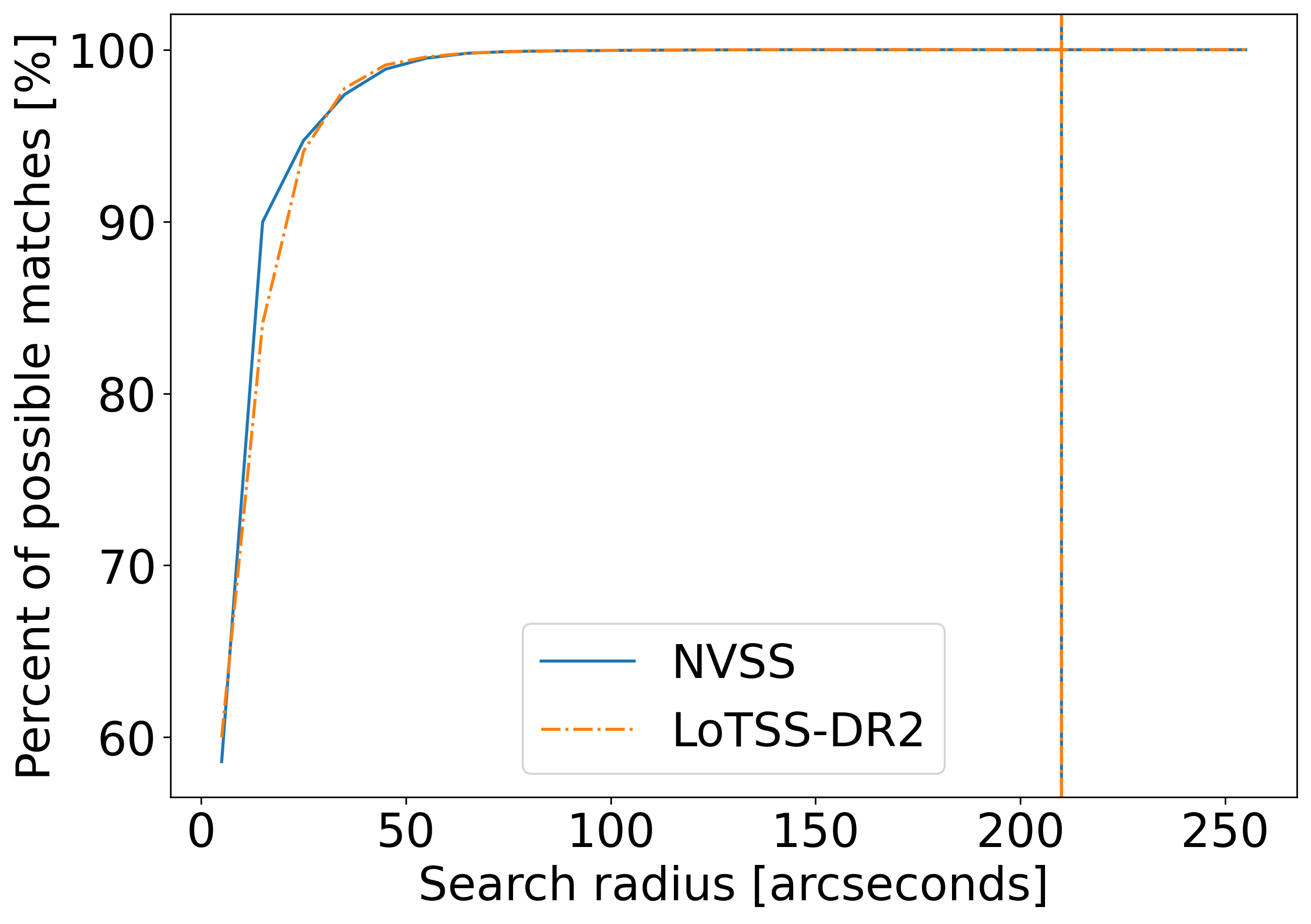}}
    \caption{Relation between search radius and included percentage of accepted matches. The dotted vertical lines give the radius of 100\% completeness. The 100\% completeness was calculated by finding all accepted matches in a $1\,000''$ search radius.}
    \label{fig:search_radius}
\end{figure}

If the potential counterpart was not accepted, in a second step another value is calculated. For this, both source centres are connected with a line and $v$ is calculated at the intersection with the FWHM ellipse of the counterpart source (red dots in Fig.~\ref{fig:matching}). If $v$ at one of these coordinates is above $1.1 v_{thr} = 0.55$, the pair is also accepted. The value of $1.1$ was chosen from testing the algorithm in several example situations and lead to more consistent matching results. If it were chosen as $1.0$, it would be enough for the two FWHM to only touch at one point, but as some overlap is demanded for physical matches, this case is excluded. If the value were chosen higher, at $1.2$, the condition would be too strict and matches missing. This value may offer some room for improvement in further iterations of the method, but was out of the scope for this first release.
This additional property ensures the correct identification of elongated sources while the slightly higher threshold keeps misclassifications low. An example of this second step is given in Fig.~\ref{fig:matching} (the upper left orange contour).
Here $205''$ was chosen as the search radius, as it is the lowest radius which includes all accepted matches which were found up to $1\,000''$. This is demonstrated in Fig.~\ref{fig:search_radius}.


\subsection{Artefacts and cleaning}

\begin{figure}
    \resizebox{\hsize}{!}{\includegraphics{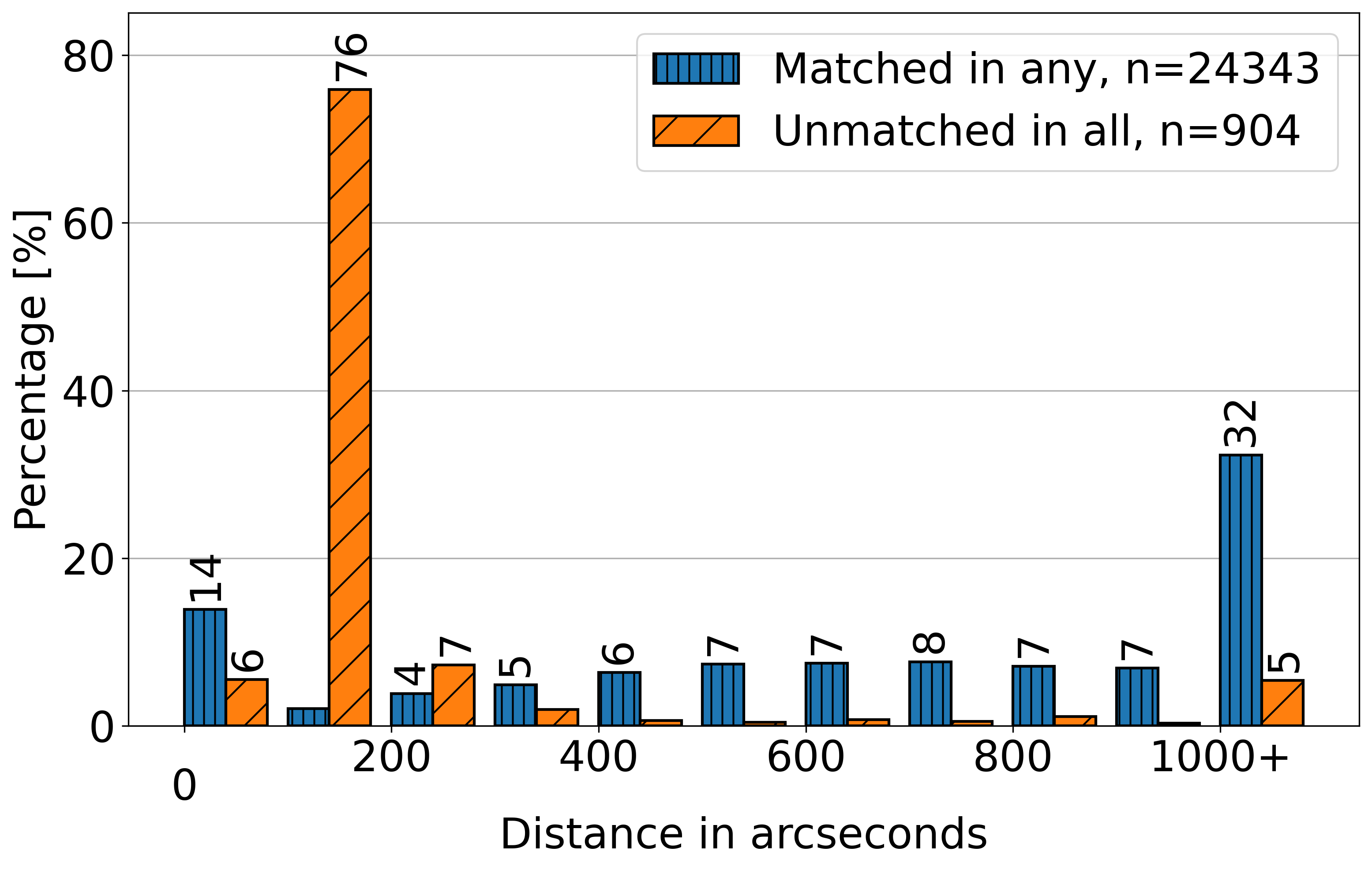}}
    \caption{The percentage of sources for a given distance from the nearest bright sources -- defined here as brighter than 0.5 Jy in LoLSS. The 1\,000+ bin includes all sources with a distance greater than $1\,000''$ from the nearest bright source and therefore has a different area than the other bins.}
    \label{fig:artefact_unmatched}
\end{figure}

To refine these results, the unmatched LoLSS sources are checked for their distance to bright sources, as the noise is especially high in the vicinity of bright sources. This noise may have been picked up by the source finder PyBDSF and imaging artefacts were thereby added to the source catalogue. In Fig.~\ref{fig:artefact_unmatched} the distance of matched (in any survey) and unmatched (in all surveys) sources to the next bright source ($>0.5$~Jy in LoLSS) is displayed. For the matched sources, a nearly constant count above $400''$ is apparent. For the unmatched sources, around 75\% of the sources are closer than $200''$ to a bright source and 85\% closer than $300''$. Since above $300''$ a constant relation can be seen, all unmatched sources (no match in any survey) up to $300''$ close to a bright source are removed. This labels about 819 LoLSS sources as artefacts. 

For the remaining unmatched sources, a visual inspection was done which led to the conclusion that at least 90\% of these are artefacts and not actual sources. This is apparent due to their closeness to sources with calibration errors and the comparison with the higher resolution survey LoTSS-DR1 or -DR2. Therefore, as a conservative choice, all unmatched sources are removed. Overall 956 unmatched LoLSS sources have been flagged as artefacts, which corresponds to a total of 3.79\% of the LoLSS catalogue. 

In the publication accompanying the release of the data set \citep{deGasperin2021}, 1\,055 sources (4\%) were estimated as being false positives. These were found to be mostly concentrated at the edges and around bright sources. This is in good agreement with the findings presented in this work.

\begin{figure}
    \resizebox{\hsize}{!}{\includegraphics{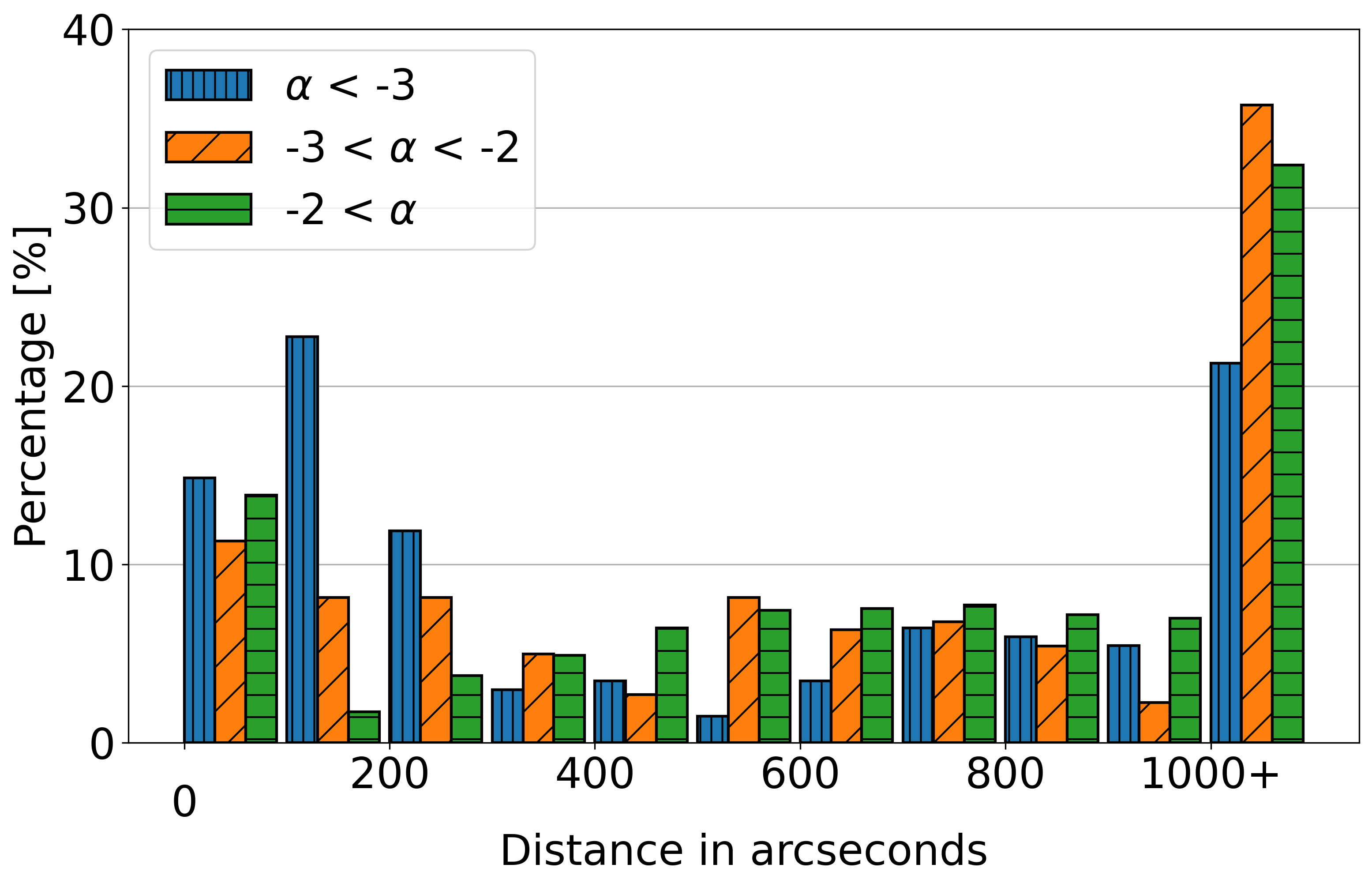}}
    \caption{The percentage of sources in three spectral index bins for a given distance from the nearest bright sources -- defined here as brighter than 0.5 Jy in LoLSS. The 1\,000+ bin includes all sources with a distance greater than $1\,000''$ from the nearest bright source and therefore has a different area than the other bins.}
    \label{fig:artefact_steep}
\end{figure}

In addition to the anomalously high number of unmatched sources close to bright sources, we also find that a high number ($\approx 40\%$) of steep ($\alpha < -3$) sources are located up to $300''$ around bright sources. In Fig.~\ref{fig:artefact_steep} the distance distribution of sources in three spectral index bins is shown. Combined with the previous removal of unmatched sources below $300''$ around bright sources, we decide to make a general cut and remove all sources closer than $300''$ to any bright source. These are labelled as artefacts in the catalogue although their origin is still unclear. Despite labelling them as artefacts, some of those sources might nevertheless be real, but may be recovered with largely offset flux density, especially in LoLSS-PR due to the missing direction-dependent calibration. The offset could be due to the presence of a bright source in the neighbourhood leading to imaging artefacts, while a higher LoLSS flux density leads to a steeper spectral index. With this additional 6\%, an overall 2\,640 (10.5\%) sources are labelled as artefacts.

\subsection{Properties of the catalogue}\label{subsec:Props}

\begin{figure}
    \resizebox{\hsize}{!}{\includegraphics{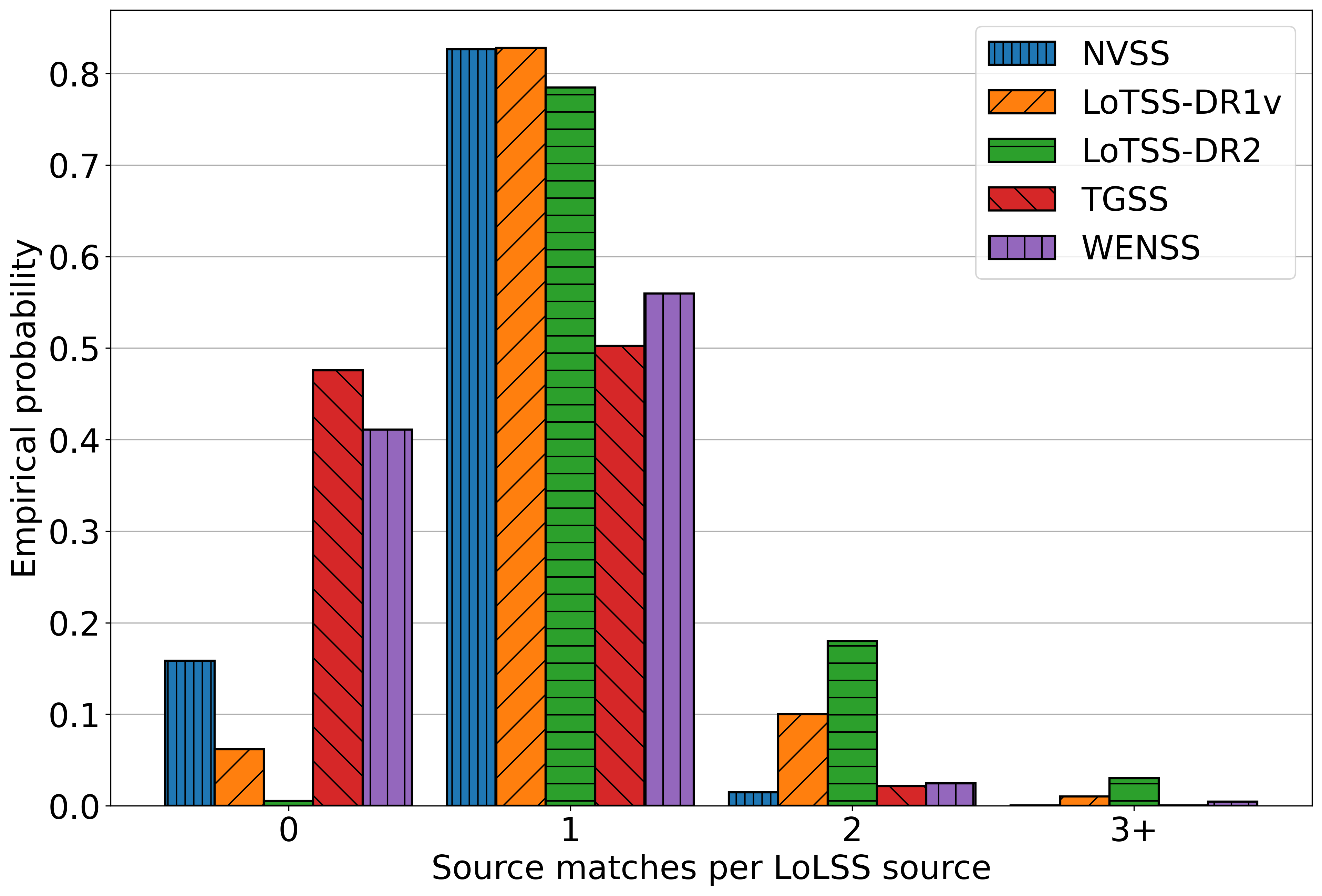}}
    \caption{The empirical probability for the number of source matches for LoTSS-DR1 value-added, LoTSS-DR2, NVSS, TGSS and WENSS per non-artefact LoLSS source.}
    \label{fig:cross_match}
\end{figure}

The cross-matching is done pairwise and the base catalogue is in all cases LoLSS. It is cross-matched with LoTSS-DR1, LoTSS-DR2, TGSS, WENSS and NVSS. The final catalogue is described in Table \ref{tab:catalogue}.
The number of matches per LoLSS source for each counterpart survey is shown in Fig.~\ref{fig:cross_match}. There it is apparent that most sources have exactly one counterpart and LoTSS-DR2 has the most matches. Only very few sources $(<150)$ do not have a counterpart in LoTSS-DR2, but in other surveys. These are mostly due to problems with the source finding in LoLSS-PR, where one ellipse is ill fitted to two lobes without a core. This should be fixed in further iterations of the LoLSS survey. 
Few LoLSS sources -- less than 20\% -- have two counterparts in LoTSS-DR2. In the other catalogues very few sources ($<10$\%) have more than one counterpart. For NVSS with a low source density and similar resolution to LoLSS, less than 300 multiple counterparts (2+) are found, while for 80\% of the LoLSS sources a one-to-one match was found. In TGSS and WENSS, no counterpart was found for around 45\% of the LoLSS sources. Our analysis does not further investigate whether multiple associations are exclusively physical or contain chance associations due to blending.

The cross-matching catalogue has 25\,247 entries and is available at CDS, of these 2\,640 have no counterpart and are labelled as 'artefacts' (A). Of the remaining, 16\,266 are labelled as 'single' (S) -- when all matches are singular -- and 6\,341 as 'multiple' (M), if in any survey multiple counterparts were matched. 

\begin{figure}
    \resizebox{\hsize}{!}{\includegraphics{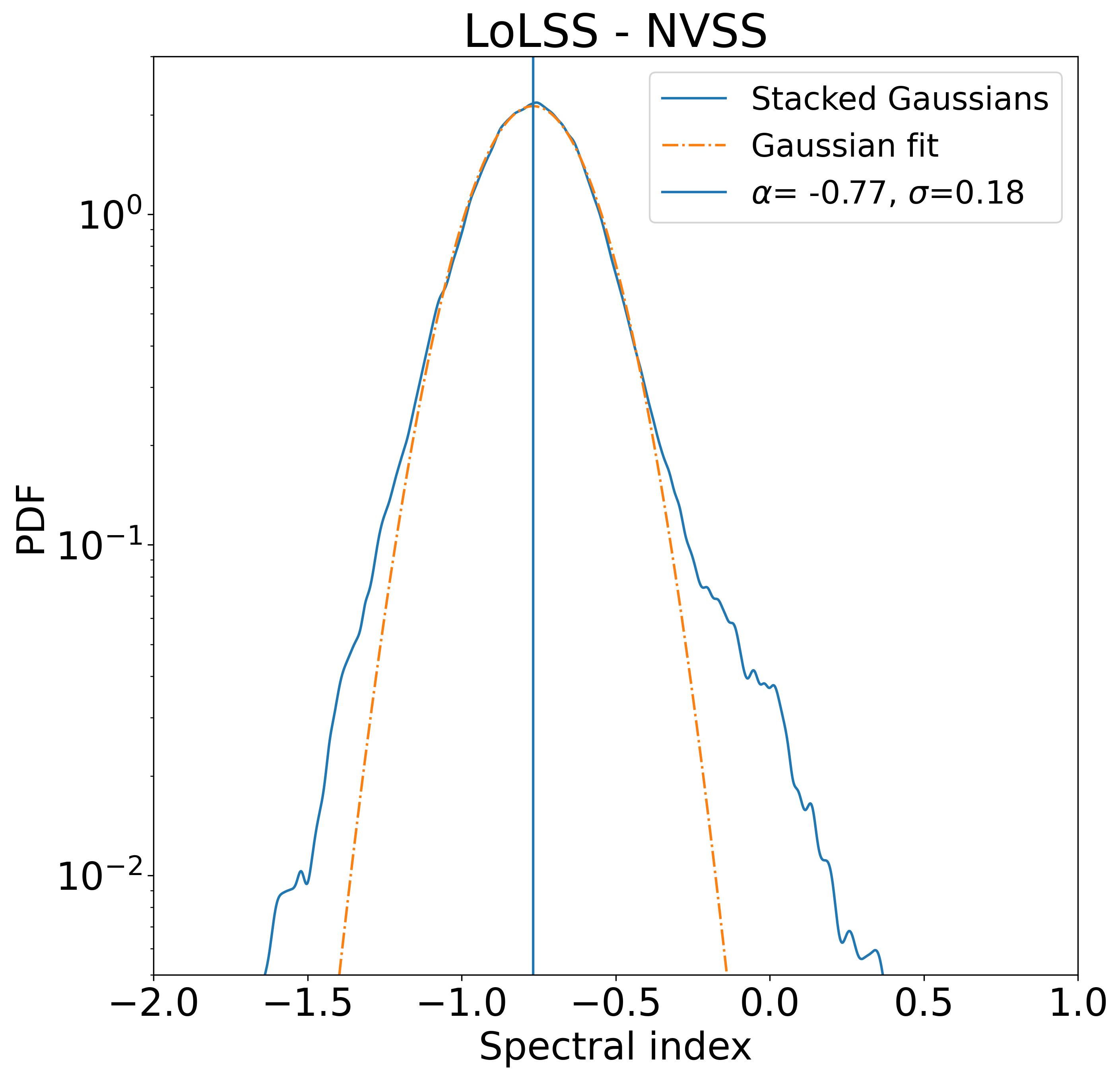}}
    \caption{Spectral index distribution for sources matched in LoLSS and NVSS. The orange line is a Gaussian fit with the parameters noted in the legend.}
    \label{fig:alpha_LN}
\end{figure}

\begin{figure}
    \resizebox{0.975\hsize}{!}{\includegraphics{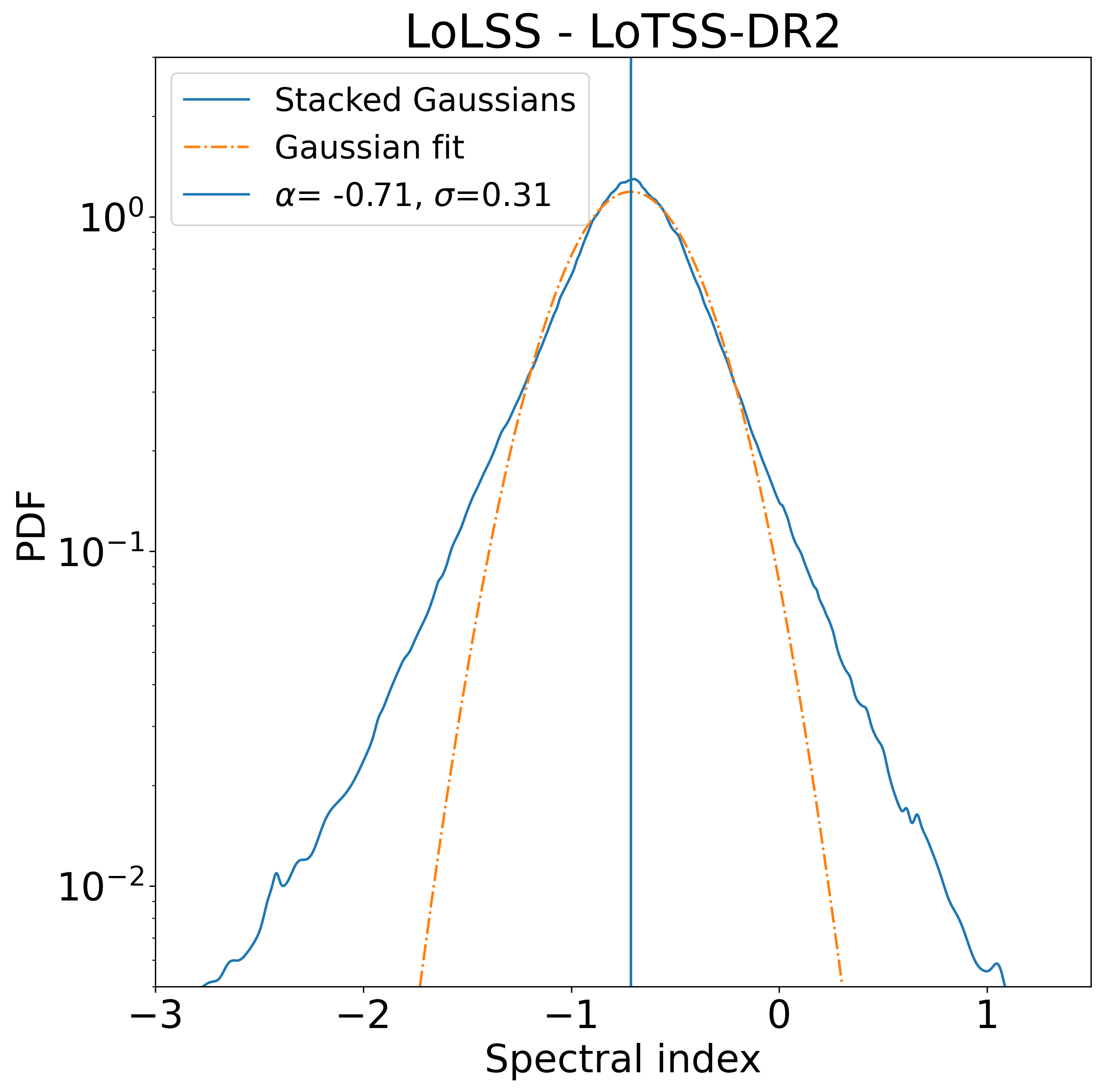}}
    \caption{Spectral index distribution for sources matched in LoLSS and LoTSS-DR2. The orange line is a Gaussian fit with the parameters noted in the legend.}
    \label{fig:alpha_LT2}
\end{figure}

For all matches, the spectral index is calculated by bootstrapping 1\,000 times the flux density values for each survey from a Gaussian distribution. For sources of type S, it is centred on the measured flux density value with a standard deviation according to the flux density error given in the catalogue. For sources of type M, the Gaussian distribution is centred on the sum of the individual flux densities, while the error is given by the root mean square of the individual errors.
If no flux density error is given, 10\% of the flux density is used. The resulting mean is used as the spectral index and the standard deviation as the corresponding error. A chance association to any given association normally has a very small impact on the spectral index, as in most of these cases the real association will be significantly brighter.
This method is the same as \citet{degasperin2018} used to calculate the spectral index of TGSS--NVSS. The normalised spectral index distribution is then retrieved by stacking normalised Gaussians with the values retrieved from the bootstrap and dividing by the amount of spectral indices. The spectral index distributions for LoLSS--NVSS and LoLSS--LoTSS-DR2 are shown in Fig.~\ref{fig:alpha_LN} and Fig.~\ref{fig:alpha_LT2}, respectively. For the LoLSS--NVSS spectral indices a narrow peak at $-0.77\pm0.18$ is found after fitting a Gaussian to the central distribution. It is also apparent that the distribution is skewed towards positive values. 
For the LoLSS--LoTSS-DR2 a broader peak centred at $-0.71\pm0.31$ is found, with significant counts down to $-3$ and up to $1$. 
To check whether the multiple matches are physical matches or just chance alignments, we also compared the spectral index of the single and multiple matches. Both distributions only differ by $\Delta \alpha = 0.002$ and $\Delta \sigma = 0.03$.
%
The spectral index distribution of LoLSS--TGSS and LoLSS--WENSS is shown in Fig.~\ref{fig:Ap_Alpha_LG} and Fig.~\ref{fig:Ap_Alpha_LW}. These have a fitted mean spectral index of $-0.82\pm0.36$ and $-0.77\pm0.24$, for TGSS and WENSS, respectively. 

\begin{figure}
    \resizebox{\hsize}{!}{\includegraphics{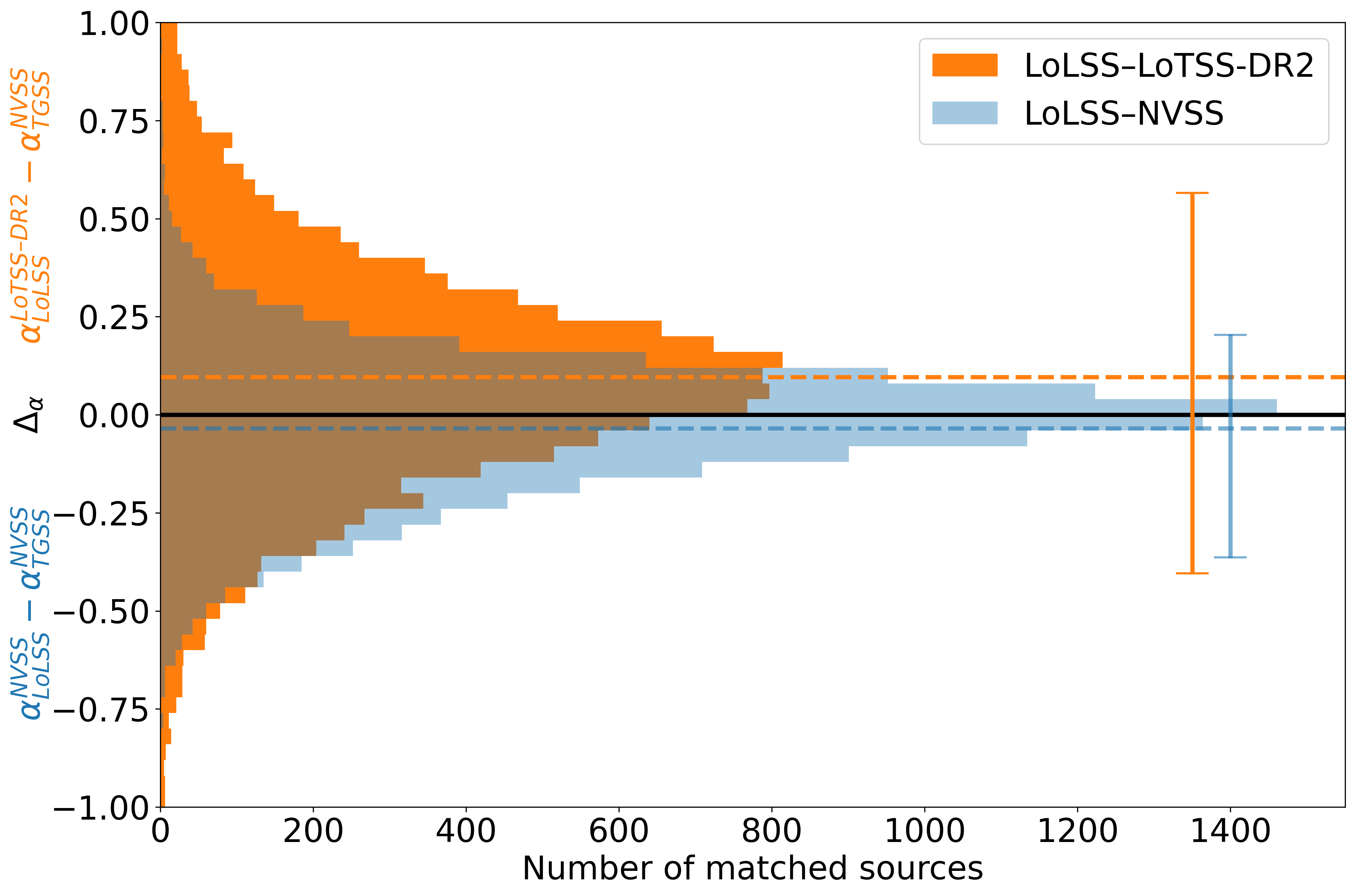}}
    \caption{Spectral index difference between two spectral indices calculated in this study (LoLSS--LoTSS-DR2 and LoLSS--NVSS) and the TGSS--NVSS spectral index catalogue of \citet{degasperin2018}. The dashed lines indicate the mean value and the bars on the right depict the 90 per cent interval of $\Delta_{\alpha}$.}
    \label{fig:alpha_diff}
\end{figure}

We compare the spectral index values from this study with a similar, previous study by \citet{degasperin2018}. There, a spectral index catalogue was retrieved from NVSS and TGSS by convolving TGSS to $45''$, re-gridding the mosaic images and running PyBDSF on these. The comparison was done by a quick positional cross-match of LoLSS source positions in a standard search radius of $5''$. For each match, the spectral index difference $\Delta_{\alpha} = \alpha_{\mathrm{LoLSS}}^{\mathrm{NVSS\ or\ LoTSS}} - \alpha_{\mathrm{TGSS}}^{\mathrm{NVSS}}$ as the difference of both spectral indices is calculated. We do this for both the LoLSS--NVSS and the LoLSS--LoTSS-DR2 matching. For NVSS we find an average $\Delta_{\alpha} = -0.034 \pm 0.178$ and for LoTSS-DR2 $\Delta_{\alpha} = 0.097 \pm 0.333$. This indicates that the LoLSS--LoTSS-DR2 spectral indices are on average flatter. In Fig.~\ref{fig:alpha_diff} the 90 per cent errors of the $\Delta_{\alpha}$ distribution are plotted.

\begin{table*} \ra{1.3}
\centering 
\caption{Description of catalogue's columns, available at CDS.}
\label{tab:catalogue}
\rowcolors{2}{lightgray!40}{}
\begin{tabular}{l l l l} \toprule
ID & Column name & Unit & Notes \\ \midrule
0 & Source{\_}id{\_}L & - & LoLSS source identification number\\
1 & RA & degree & LoLSS position\\
2 & Dec & degree & LoLSS position\\
3 & Flux{\_}LoLSS & mJy & Sum of detection fluxes \\
4 & E{\_}Flux{\_}LoLSS & mJy & Sum of detection flux errors \\
5 & Peak{\_}Flux{\_}LoLSS & mJy beam$^{-1}$ & Maximum of detection peak fluxes \\
6 & E{\_}Peak{\_}Flux{\_}LoLSS & mJy beam$^{-1}$ & Maximum of detection peak flux errors\\
7 & Rms{\_}LoLSS & mJy beam$^{-1}$ & Local rms noise\\
8--33 & & & 3--7 Repeated for all other surveys \\
33 & Spidx{\_}LT1 & - & Spectral index between LoLSS and LoTSS-DR1\\
34 & Spidx{\_}LT2 & - & Spectral index between LoLSS and LoTSS-DR2\\
35 & Spidx{\_}LG & - & Spectral index between LoLSS and TGSS\\
36 & Spidx{\_}LW & - & Spectral index between LoLSS and WENSS\\
37 & Spidx{\_}LN & - & Spectral index between LoLSS and NVSS\\
38 & Scode & S,M,A & S = Single, M = Multiple, A = Artefact/Close to bright source \\
39 & Num{\_}match{\_}T1 & - & Number of LoTSS-DR1 counterparts \\
40 & Source{\_}id{\_}T1 & - & Array of source IDs with Num{\_}match{\_}T1 entries \\
41 & Num{\_}match{\_}T2 & - & Number of LoTSS-DR2 counterparts\\
42 & Source{\_}id{\_}T2 & - & Array with Num{\_}match{\_}T2 entries\\
43 & Num{\_}match{\_}G & - & Number of TGSS counterparts\\
44 & Source{\_}id{\_}G & - & Array with Num{\_}match{\_}G entries\\
45 & Num{\_}match{\_}W & - & Number of WENSS counterparts\\
46 & Source{\_}id{\_}W & - & Array with Num{\_}match{\_}W entries\\
47 & Num{\_}match{\_}N & - & Number of NVSS counterparts\\
48 & Source{\_}id{\_}N & - & Array with Num{\_}match{\_}N entries\\\bottomrule
\end{tabular} 
\end{table*}

\subsection{Cross-matching incompleteness}

The cross-matching incompleteness can be estimated for each survey matched with LoLSS by the number of unmatched, non-artefact LoLSS sources. For this, LoLSS is divided into logarithmic flux density bins and in each bin the fraction of unmatched to the total number of LoLSS sources in that bin is calculated. 
This incompleteness is shown in the right panel of Fig.~\ref{fig:alpha_LT2_flux} for LoTSS-DR2. At low LoLSS flux densities, the incompleteness is very low at the $10^{-3}$ level, indicating that virtually all sources have a counterpart in LoTSS-DR2. Overall the incompleteness stays around the 1 per cent level. Therefore this cross-matching is always $\geq 99$ per cent cross-matching complete. 
The incompleteness of the cross-matching for the other surveys with LoLSS is shown in Fig.~\ref{fig:Ap_Incomplete}. A strong similarity between NVSS and TGSS can be seen. Both are close to an incompleteness of one at very faint flux densities and only reach around $2\times 10^{-2}$ as the lowest incompleteness. At high flux densities both become more incomplete, though this is biased due to less populated bins at high flux densities. 
At low and mid flux densities the WENSS match behaves similar, reaches values below $10^{-2}$ and is even 100\% complete above $0.5 \log_{10}$ Jy.


\subsection{Specific sub-catalogues}\label{subsec:cat}
As the whole catalogue is rather large and sources for further scientific research may not be immediately visible, we produce three additional, smaller sub-catalogues -- also available at CDS -- which highlight sources with different characteristics, though all are candidates for high-redshift radio galaxies (HzRGs). Many HzRGs are hosted by some of the most luminous and massive galaxies, which may be the predecessors for massive ellipticals at low redshifts \citep{best1998,carilli2002,reuland2004,afonso2011,saxena2019}. One successful tracer for HzRGs is the relation between radio spectra and redshift, as a large fraction of the steepest spectra sources also have a high-redshift \citep{miley2008}.

These sub-catalogues are all based on the LoLSS--LoTSS-DR2 cross-matching.
The first subcatalogue is based solely on ultra steep spectrum (USS) sources and includes 292 sources with a spectral index below minus two and may still include artefacts.
The second one includes 9 high-redshift ($z>2$) and steep spectrum ($\alpha < -1$) sources, which were visually inspected for artefacts or matching errors.
For the third subcatalogue, a stronger selection criteria based on \citet{saxena2019} is chosen: compact (MAJ < $10''$ \& Peak/Total flux density > 0.8), single component and steep spectrum sources ($\alpha_{54}^{144} < -1.3$) resulting in 46 sources. All conditions are evaluated in LoTSS-DR2 and the sources visually inspected. All 46 sources have a flux density below $S_{144} = 50$ mJy, which was the lowest flux density in the sample of \citet{saxena2019}, therefore no overlap is expected.



\section{Discussion}\label{Sec4}
\subsection{Cross-matching method}
We compare our cross-matching method also with the simple positional cross-matching. For this, we try two different matching radii, $50''$ – similar to the resolution of LoLSS – and $12''$, two times the resolution of LoTSS. For a radius of $50''$, it leads to more multiple matches while also resulting in a more enhanced tail of steep spectral sources, therefore indicating that these are chance associations (random matches lead to steep indices). When using a radius of $12''$ for the positional cross-matching, significantly less counterparts are matched and most doubles in LoTSS-DR2 are missing. Therefore we conclude that our algorithm provides better results than a simple positional cross-matching.

The other introduced cross-matching methods from Sec.~\ref{Sec1}, which were developed for cross-identifying radio sources in optical surveys should also perform worse than our developed method. That is because they can only take into account the source extensions from one survey, while the method introduced in this paper uses the source extensions from both surveys. This method should also be able to do the optical cross-identification between extended (radio) and point-like (optical) sources, though it will perform worse than for example \cite{fan2020}, which will in addition also cross-match the lobes and core at the same time. However this is not the problem for which our method was developed, our goal was to match extended sources with extended sources from different radio surveys. For this case it can also be applied to all other radio surveys, or even in other cases where a matching of extended/resolved sources is needed.

\subsection{Spectral index and flux density}
Previous studies have found conflicting results as to how the median spectral index changes with flux density. Some found a flatter spectral index at low flux densities \citep{prandoni2006,ishwara-chandra2010,intema2011,williams2021}, though with inconsistencies in the rate of change between them. In other studies a constant spectral index was observed \citep{kapahi1986,sirothia2009}.

\begin{figure}
    \resizebox{\hsize}{!}{\includegraphics{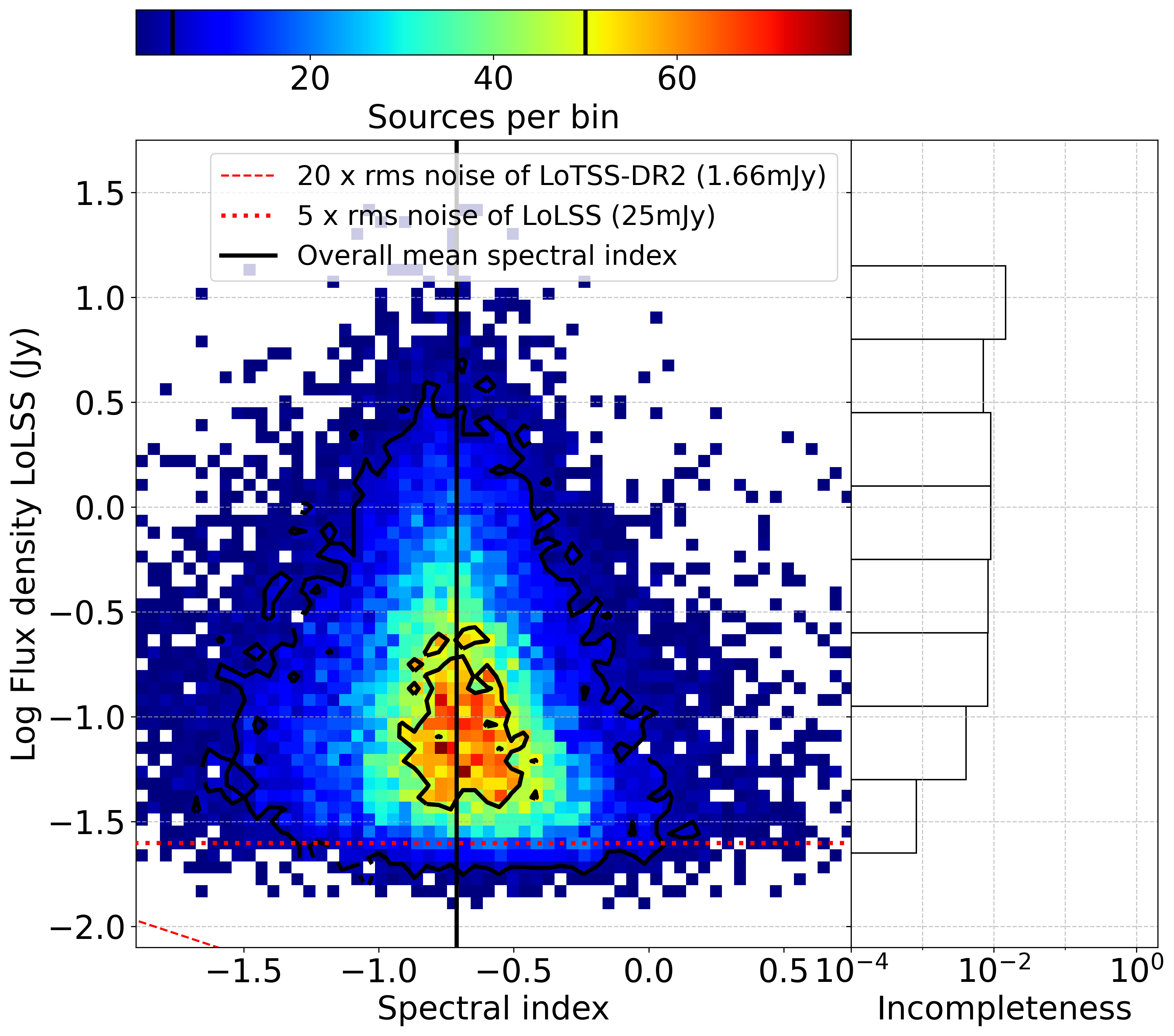}}
    \caption{Spectral index and flux density 2d-histogram. Right plot gives the incompleteness as the ratio of unmatched to all sources in a given flux density bin. The red dashed lines indicate the flux density limits of the surveys, while the black vertical line indicates the average spectral index. The black lines in the color bar define the contours.}
    \label{fig:alpha_LT2_flux}
\end{figure}
Figure \ref{fig:alpha_LT2_flux} shows the spectral indices between LoLSS and LoTSS-DR2 along with the LoLSS flux densities in the main panel and the above-mentioned matching incompleteness in the right panel. 
The two red dashed lines indicate two flux density values: $5 \times$ rms noise of LoLSS ($25$ mJy) and $20 \times$ rms noise of LoTSS-DR2 ($1.66$ mJy). The LoTSS-DR2 value was chosen with a factor of 20, as it would not show up otherwise due to being too low. The vertical black line indicates the average spectral index as found before in Fig.~\ref{fig:alpha_LT2}. Above $5 \times$ the rms noise of LoLSS, no bias due to incomplete data impacts the results.
Overall the majority of sources can be found at the fainter flux density bins. The whole shape appears nearly symmetric around the mean spectral index.
For the right panel, the cross-matching incompleteness is calculated as explained before bin-wise for ten bins between -2 and 1.5 in units of $(\log_{10}$ Jy). The first and last bin do not show up on the histogram, as their incompleteness is zero, and they can also not be trusted, as their counts are too low.



\begin{figure}
    \resizebox{\hsize}{!}{\includegraphics{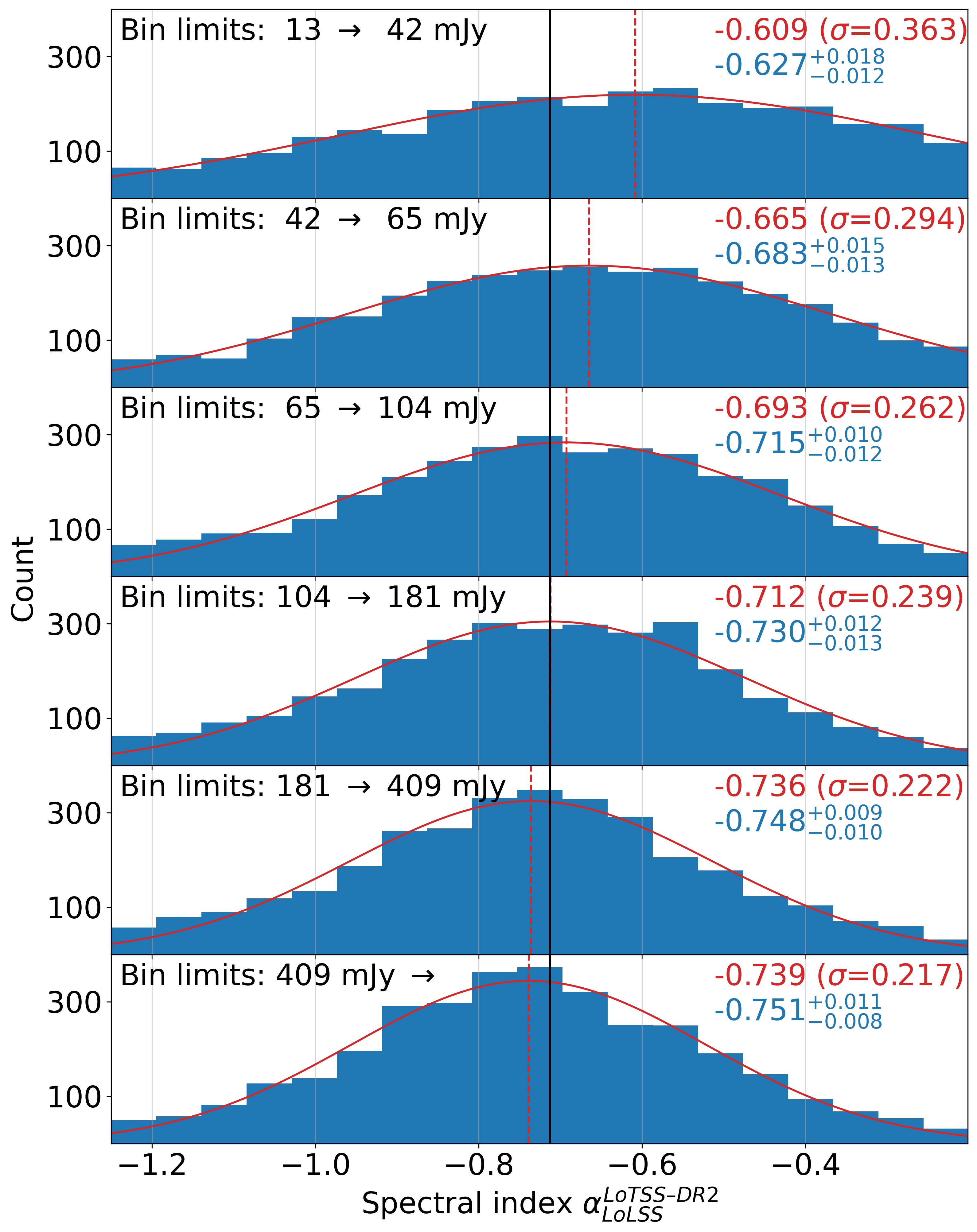}}
    \caption{Spectral index $\alpha_{\mathrm{LoLSS}}^{\mathrm{LoTSS-DR2}}$ distribution in six equally-numbered flux density bins at 54 MHz. In each bin a Gaussian fit and its centre is marked and the fit details (mean and standard deviation) are in red in the top right corner. Below that in blue is the median along with the 95\% confidence interval ranges. In each top left corner the flux density limits of the bin is noted, while the black line indicates the overall mean as previously found in Fig.~\ref{fig:alpha_LT2}.}
    \label{fig:alpha_LT2_flux_bin}
\end{figure}

For a closer inspection of the flux density dependence 
of the spectral index, the spectral index is binned in six 
equally-populated flux density bins and the histogram is 
shown in Fig.~\ref{fig:alpha_LT2_flux_bin}. For each bin a 
Gaussian is fit to the data, to retrieve the Gaussian mean 
and standard deviation, additionally the sample 
median is calculated. Since LoTSS-DR2 has much better 
sensitivity than LoLSS, no limitation on the spectral index 
is expected, even the faintest flux density bin should be 
complete above $\alpha = -3.51$. A flatter spectral index
can be seen in the three faintest bins, where the distributions 
are also more widespread ($\sigma=0.36$ in the first bin, 
$\sigma=0.26$ in the third bin). Above a flux density of 
$S_{54}=181$ mJy the mean spectral index is found to be independent of 
flux density.




\begin{figure}
    \resizebox{\hsize}{!}{\includegraphics{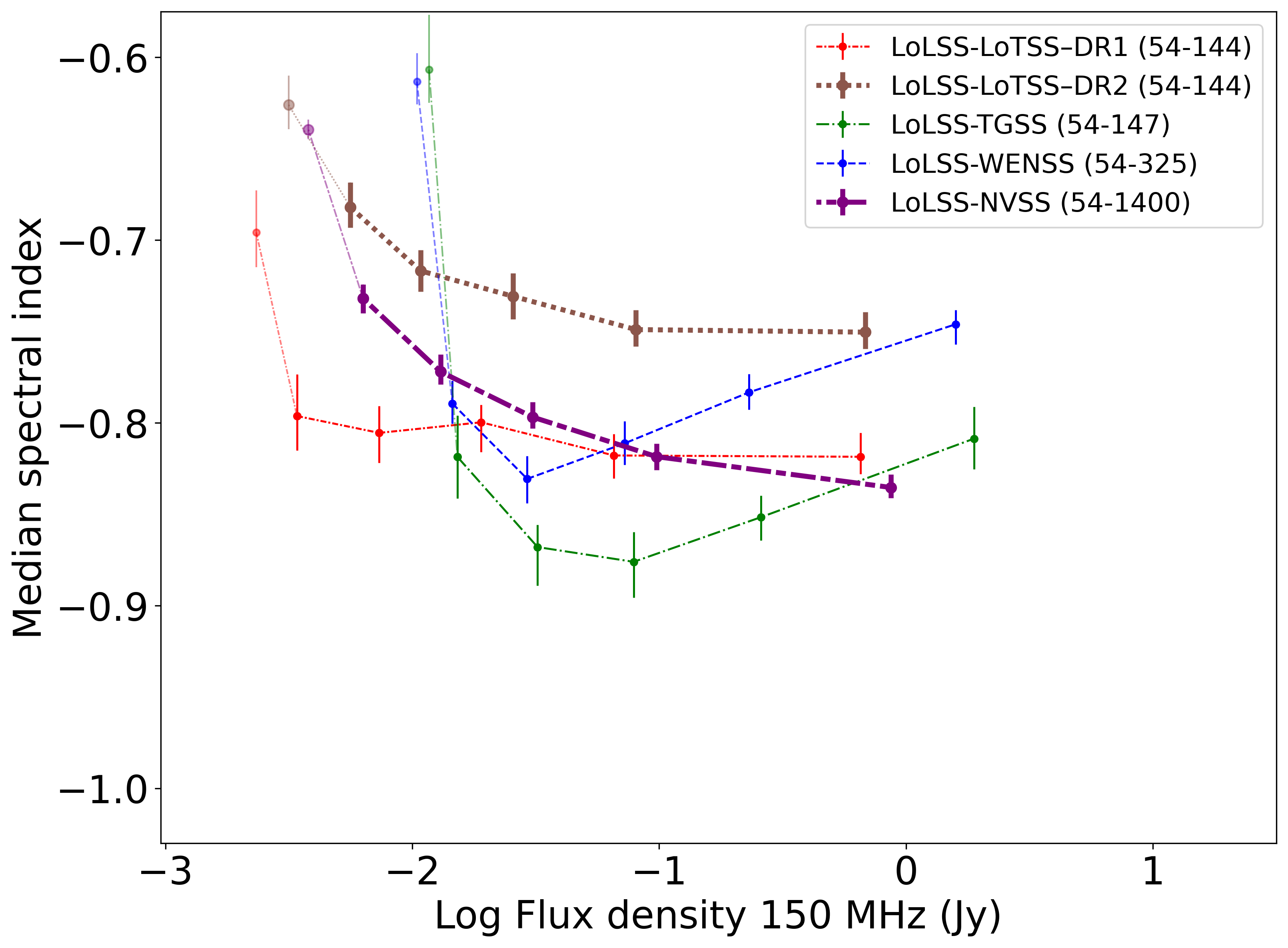}}
    \caption{Comparison of median spectral index in flux density bins for the surveys introduced in Sect.~\ref{Sec2}. On the x-axis the LoLSS flux density, scaled to $150$ MHz according to the spectral index of each bin, is shown. The 95\% confidence intervals are shown by the error bars. The first bin is grayed out due to completeness concerns.}
    \label{fig:alpha_my_comp}
\end{figure}
\begin{figure}
    \resizebox{\hsize}{!}{\includegraphics{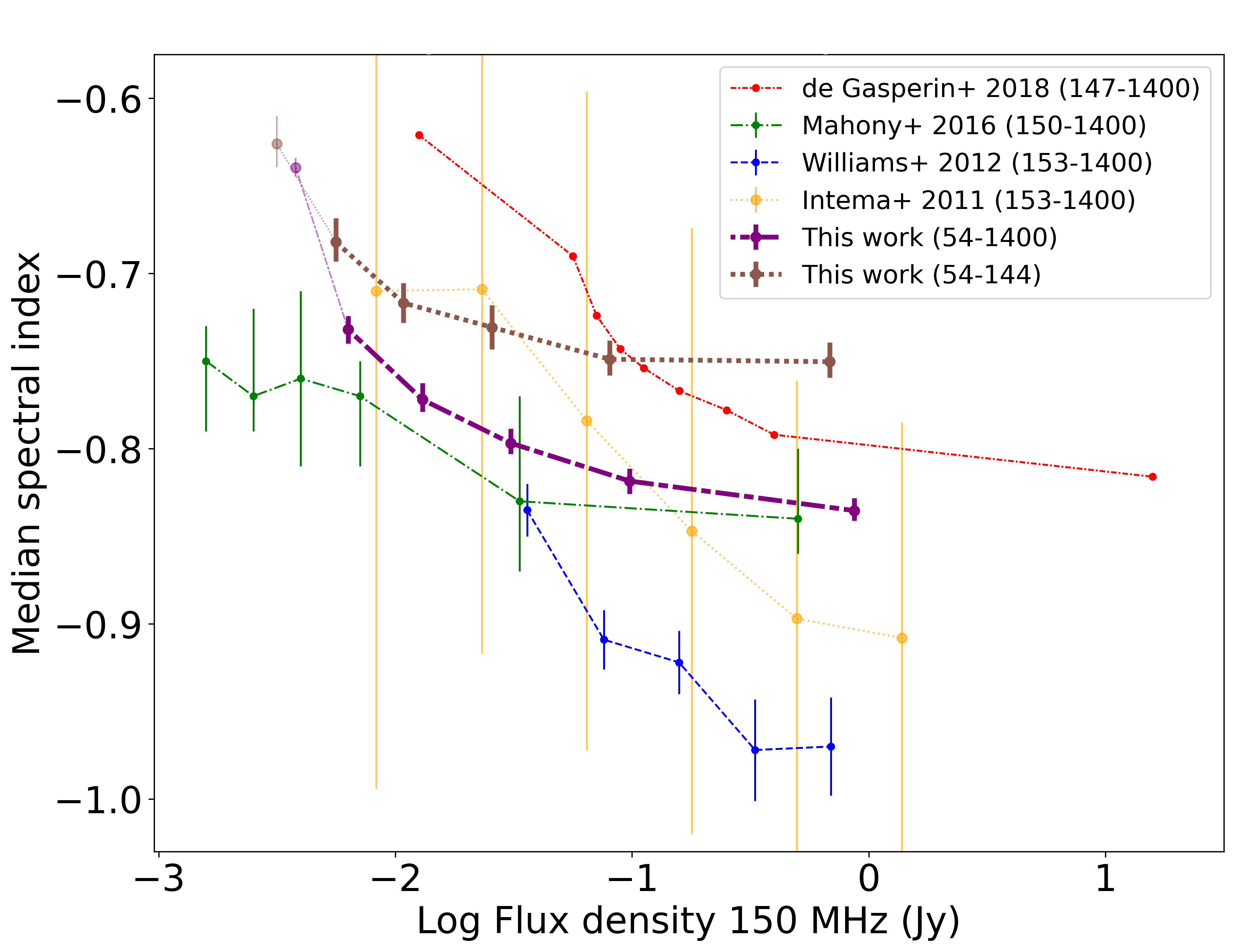}}
    \caption{Comparison of median spectral index in flux density bins. Data is taken from \citet{intema2011,williams2013,mahony2016,degasperin2018} and Fig.~\ref{fig:alpha_LT2_flux_bin} of this work. Data from this work is scaled to $150$ MHz according to the spectral index of each bin. The 95\% confidence intervals are shown by the error bars for this work. The first bin for each line from this work is greyed out due to completeness concerns.}
    \label{fig:alpha_comp}
\end{figure}

In the following, the median spectral index is compared between the different surveys used in this study. In Fig.~\ref{fig:alpha_my_comp} the median spectral index between LoLSS and each survey is shown. For this, as before, the median spectral index is calculated in six equally-populated flux density bins. Since no flux density cuts are applied to any survey catalogue, the first bin may feature incomplete data and is therefore not comparable and greyed out. Overall, for each survey pair a flattening in at least the lowest flux density bin is apparent, though sometimes within error bars. The LoLSS-LoTSS–DR1 spectral index is the only one for which all bins but the first (which may be incomplete) show a near constant value. It is important to note that due to an improved flux density calibration scheme for LoTSS–DR2, which is aligned with NVSS and 6C, the values differ on the order of up to $10\%$ to LoTSS–DR1. Therefore the spectral indices involving these two are different. For the TGSS and WENSS match a similar shape is found, where both spectral indices become flatter again with higher flux density, but the TGSS shows more negative values in all except the first bin. 
Remaining discrepancies between the various results may either be due to systematic differences in the flux density scale on the order of up to 10 per cent in the different surveys (TGSS, WENSS, LoTSS–DR1 compared to LoTSS–DR2 and NVSS), which can lead to a difference in the spectral index of up to $0.1$, but should not influence trends within each result. However, for LoLSS-LoTSS–DR2 and LoLSS-NVSS, which are on the same flux density scale, the differences may be due to spectral curvature, see Sect.~\ref{subsec:spec_curv}.

The median spectral index is also compared to similar studies in the frequency range of 150 to 1\,400 MHz by \citet{intema2011,williams2013,mahony2016,degasperin2018} in Fig.~\ref{fig:alpha_comp}. We compare these to our calculation of the spectral index between LoLSS and LoTSS-DR2 and LoLSS and NVSS. In all studies, including our results, we find the same trend of a flattening of the spectral index at the lower flux densities. 
Remaining discrepancies in Fig.~\ref{fig:alpha_comp} may be due to the usage of different generations of LoTSS data, each with a different flux density calibration, and TGSS. Though the trends within each result should be consistent.


\subsubsection{Compact and extended sources}
\begin{figure}
    \resizebox{\hsize}{!}{\includegraphics{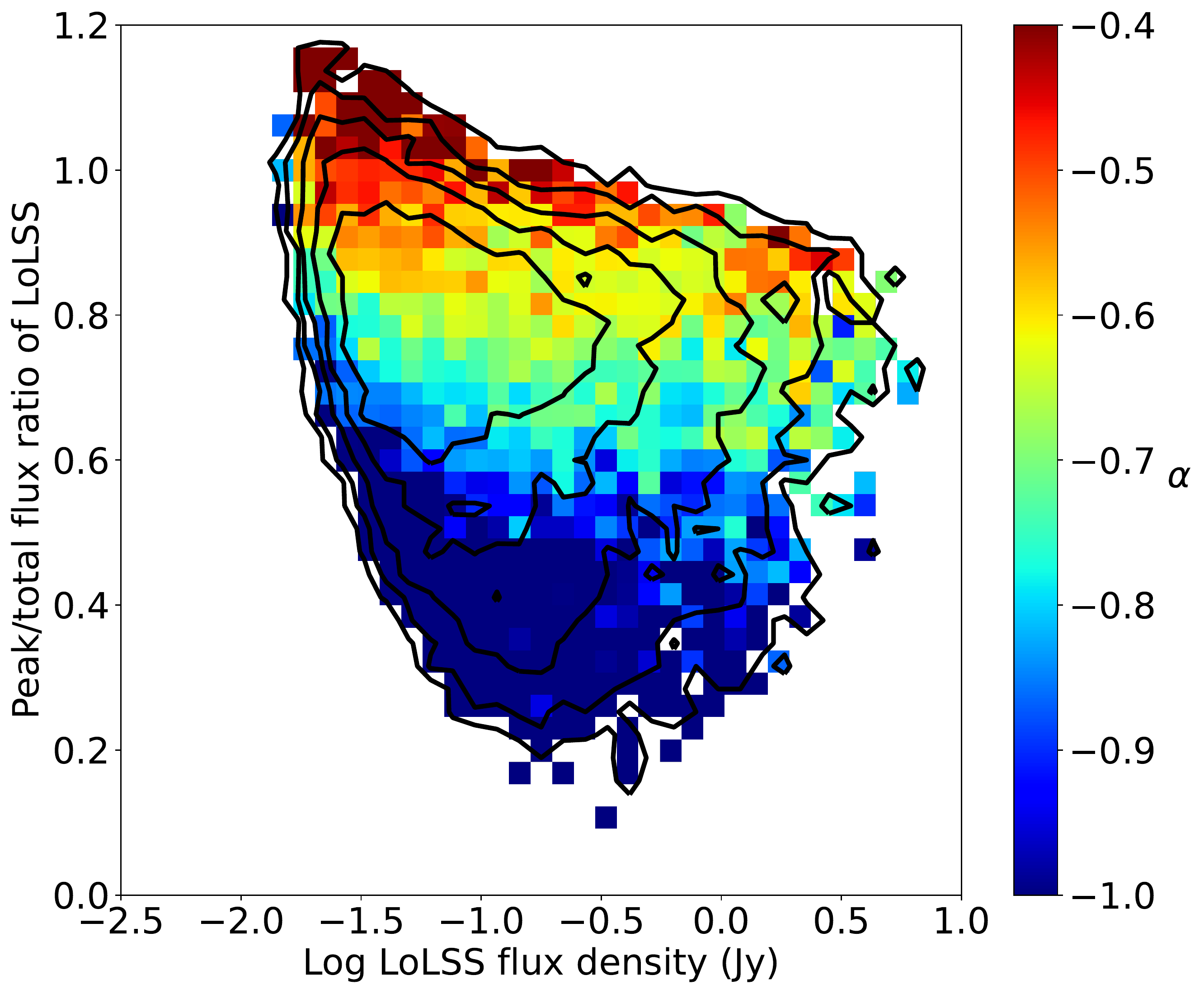}}
    \caption{LoLSS flux density plotted against peak-to-total flux ratio from LoLSS with the spectral index to LoTSS-DR2 as colour on the z-axis. The contours indicate the number counts at the 5, 10, 20, 40 and 80 levels.}
    \label{fig:PtI_L}
\end{figure}

The flux density dependence is further observed based on the compactness of sources. The compactness is calculated as the peak-to-total flux ratio of the matched sources. In Fig.~\ref{fig:PtI_L} the spectral index as a function of LoLSS flux density and compactness is plotted. The plot shows that generally compact and faint sources have a flatter spectrum than extended or bright sources. 
A possible reason for this division lies in the different parts of radio galaxy emission. Young and core-dominated radio galaxies show compact emission with a flat radio spectrum ($\alpha \approx -0.5$). In comparison, in older radio galaxies the emission of large (so less compact) lobes, which is steeper (as discussed in the introduction), dominates over the core emission, moving it more to the right (and slightly lower) in the plot. When the AGN jets shut down, the lobes become fainter, the expansion continues and the emission steepens, therefore moving more to the bottom left of the plot \citep{degasperin2018}.

\begin{figure}
    \resizebox{\hsize}{!}{\includegraphics{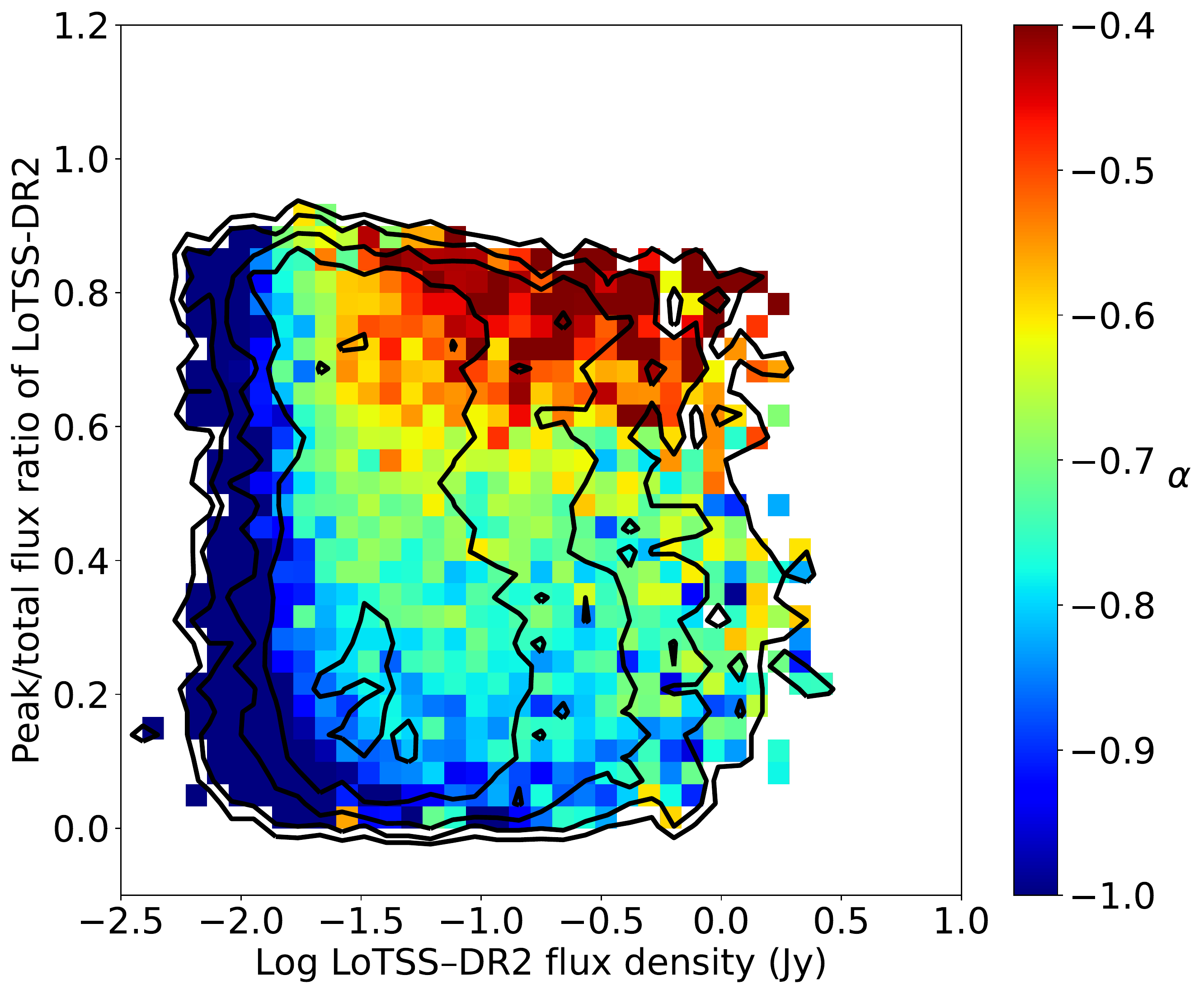}}
    \caption{LoTSS-DR2 flux density plotted against peak-to-total flux ratio from LoTSS-DR2 with the spectral index to LoLSS as colour on the z-axis. The contours indicate the number counts at the 5, 10, 20, 40 and 80 levels.}
    \label{fig:PtI_T}
\end{figure}

In Fig.~\ref{fig:PtI_T} the compactness and flux density from LoTSS-DR2 is used instead.
The difference between both plots arises from the different resolutions of $6$ and $47''$ affecting the peak flux densities, and the flux density of each bin which scales (differently) depending on its spectral index.
For the LoLSS plot Fig.~\ref{fig:PtI_L}, diffuse sources are missing due to sensitivity limitations of the LBA antennas. 


\subsection{Spectral curvature}\label{subsec:spec_curv}

\begin{figure}
    \resizebox{\hsize}{!}{\includegraphics{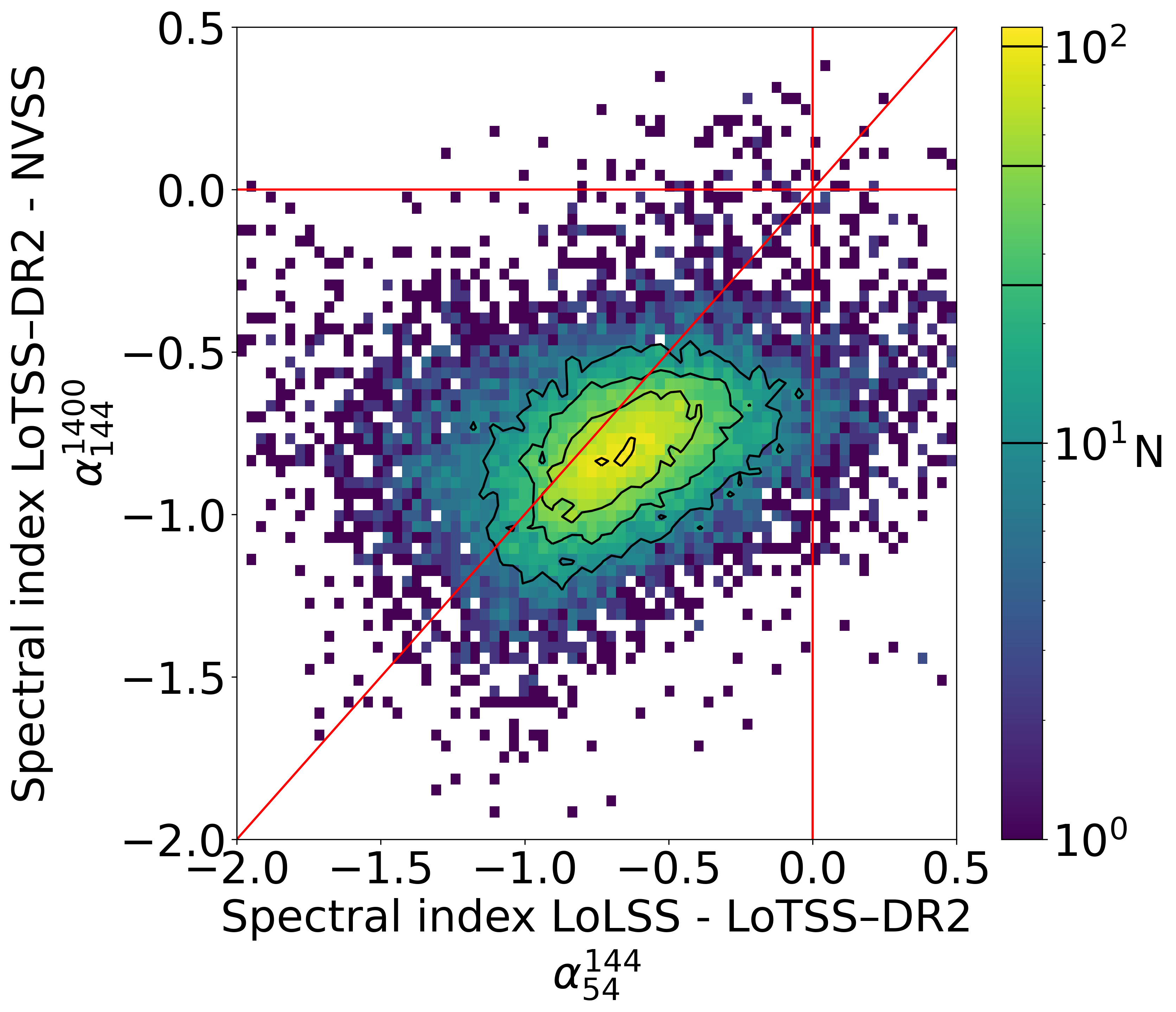}}
    \caption{Spectral curvature between LoLSS - LoTSS–DR2 - NVSS. The colour indicates the number of sources per pixel, while the contours indicate the source counts as marked on the colour scale.}
    \label{fig:Curvature}
\end{figure}
Looking at the difference of the spectral index between LoLSS--LoTSS-DR2 and LoTSS-DR2--NVSS in Fig.~\ref{fig:Curvature}, it is apparent that most sources exhibit a negatively curved spectrum, i.e. a concave shape of the SED. The mean spectral index 
for LoLSS--LoTSS-DR2 is $\alpha_{54}^{144} = -0.71 \pm 0.31$, while for LoTSS-DR2--NVSS it is found as $\alpha_{144}^{1400} = -0.80 \pm 0.17$. The peak of the 2d-histogram in Fig.~\ref{fig:Curvature} is offset from the diagonal to flatter values at the lower frequency range and steeper values for 144--1400 MHz. The difference between the spectral indices is found as $\Delta_{\alpha} = 0.089$. The same trend was already visible in Fig.~\ref{fig:alpha_my_comp}, where the spectral index between LoLSS and NVSS is flatter than LoLSS--LoTSS-DR2. This is expected because when the turnover happens at e.g. around 100 MHz, $\alpha_{54}^{144}$ becomes flatter and $\alpha_{144}^{1400}$ steeper than without flattening, while $\alpha_{54}^{1400}$ is in-between.
Due to the change to a more reliable flux density calibration for DR2, this effect became apparent. Using the LoTSS-DR1 data, no spectral curvature is evident. The observed field is big enough to rule out the effect of cosmic variance, as a test with two different RA divided samples showed.

A more detailed study of spectral curvature is possible with the direction-dependent calibration and release of LoLSS-DR1 \citep{degasperin2023}, which also include in-band spectra. In a first assessment in \cite{degasperin2023}, a similar trend in spectral curvature was found, which is even more apparent when the in-band spectra are included.

\subsection{Spectral index and redshift}

As mentioned before, the value-added LoTSS-DR1 catalogue provides (mostly photometric) redshift information for LoTSS-DR1 sources in the catalogue \citep{duncan2019}. For our subsample of LoTSS-DR2 sources matched to LoLSS, redshift information is available for 5\,662 of the 12\,491 sources.

For each survey matched with LoLSS, the spectral index above and below the mean redshift of z$_{\mathrm{mean}} = 0.736$ is almost the same, as shown in Fig.~\ref{fig:redshift}. For LoLSS--LoTSS-DR2 the difference in $\alpha$ above and below z$_{\mathrm{mean}}$ is only $0.011$, while the biggest difference is in LoLSS--TGSS with $0.035$. Which is insignificant compared to the spread of the spectral index of $0.31$. Therefore no change of the spectral index with redshift is observed. The relation was also tested for high-power radio sources ($L_{54} > 10^{27}$ W\,Hz$^{-1}$ with $z > 0.3$) with a split at $z=1$ and no difference between the two populations was found.

\section{Conclusions}\label{Sec5}
The goal of this work was to construct a suitable cross-matching method and to apply it to radio catalogues at various wavebands and angular resolutions. These are the newly released low-frequency catalogue from the LoLSS pre-release and the higher frequency catalogues from the two LoTSS data releases, TGSS, WENSS and NVSS. The development of a cross-matching method that also accounts for the different resolutions of these surveys by utilising the fitted source sizes was successful. 

In the second part of this work, a cross-matched catalogue was obtained 
and analysed. Concerning artefacts from LoLSS, \cite{deGasperin2021} estimated 1\,055 sources (4\%) to be false positives. These were found to be mostly concentrated at the edges and around bright sources. In our study we found most artefacts around bright sources and therefore used this as a criteria for artefact selection. Overall 2\,640 (10.5\%) sources are labelled as artefacts, including also sources that may be real but are likely to have unreliable flux density measurements.

Through cross-matching, the majority (16\,266 sources, 64.4\%) of LoLSS sources were found to be single component, while 6\,341 – 25.1\% – appear to be double sources. For the sources that were bright enough to be observed in NVSS, an average spectral index of 
$\alpha = -0.77 \pm 0.18$ was found. In contrast, for sources matched in LoLSS and LoTSS-DR2, a flatter average spectral index of 
$\alpha = -0.71 \pm 0.31$ was found. Comparison of the spectral slopes from LoLSS--LoTSS-DR2 with LoTSS-DR2--NVSS indicates that the probed population of radio sources exhibits evidence for a negative spectral curvature. This could also be used to extract source information like age or the spectral index of the core and lobes from the single-point spectral index measurement.

For sources matched in LoLSS and LoTSS-DR2, no flux density dependence of the spectral index above the flux density $S_{54} = 181$ mJy
was found. For fainter sources, typically a flatter spectrum is found. No evolution of the spectral index with redshift was observed.

Finally, along with the whole cross-matching catalogue, separate smaller catalogues are made available. These include different categories of sources that may be worth a deeper investigation and yield scientifically interesting results. These catalogues include 292 very steep spectrum sources, 9 high redshift, steep spectrum sources, as well as 46 potential candidate high-redshift radio galaxies (HzRGs).

\begin{acknowledgements}
LB and DJS acknowledge financial support by 
BMBF-Verbundforschung under project ErUM-Pro 
05A20PB1.
FdG acknowledges support from the Deutsche Forschungsgemeinschaft under Germany's Excellence Strategy - EXC 2121 “Quantum Universe” - 390833306.
WLW acknowledges support from the CAS-NWO programme for radio astronomy with project number 629.001.024, which is financed by the Netherlands Organisation for Scientific Research (NWO).
LOFAR data products were provided by the LOFAR Surveys Key Science project (LSKSP; \url{https://lofar-surveys.org/}) and were derived from observations with the International LOFAR Telescope (ILT). 
LOFAR \citep{LOFAR} is the Low Frequency Array designed and constructed by ASTRON. It has observing, data processing, and data storage facilities in several countries, which are owned by various parties (each with their own funding sources), and which are collectively operated by the ILT foundation under a joint scientific policy. The efforts of the LSKSP have benefited from funding from the European Research Council, NOVA, NWO, CNRS-INSU, the SURF Co-operative, the UK Science and Technology Funding Council and the Jülich Supercomputing Centre. 
The GMRT is run by the National Centre for Radio Astrophysics of the Tata Institute of Fundamental Research. 
The VLA is run by the National Radio Astronomy Observatory, a facility of the National Science Foundation operated under cooperative agreement by Associated Universities, Inc.
WENSS is a joint project of the NFRA and Leiden Observatory. 
The WSRT is operated by the NFRA with financial support from NWO.
This research made use of Astropy, a communitydeveloped core Python package for astronomy \citep{Astropy} hosted at \url{http://www.astropy.org/}, TOPCAT \citep{Topcat}, matplotlib \citep{Matplotlib}, NumPy \citep{Numpy}, lmfit \citep{Lmfit} and SciPy \citep{Scipy}. 
\end{acknowledgements}

\bibliographystyle{aa}
\bibliography{Library}

\begin{appendix}
\section{Additional Figures}
\begin{figure}[h!]
    \resizebox{\hsize}{!}{\includegraphics{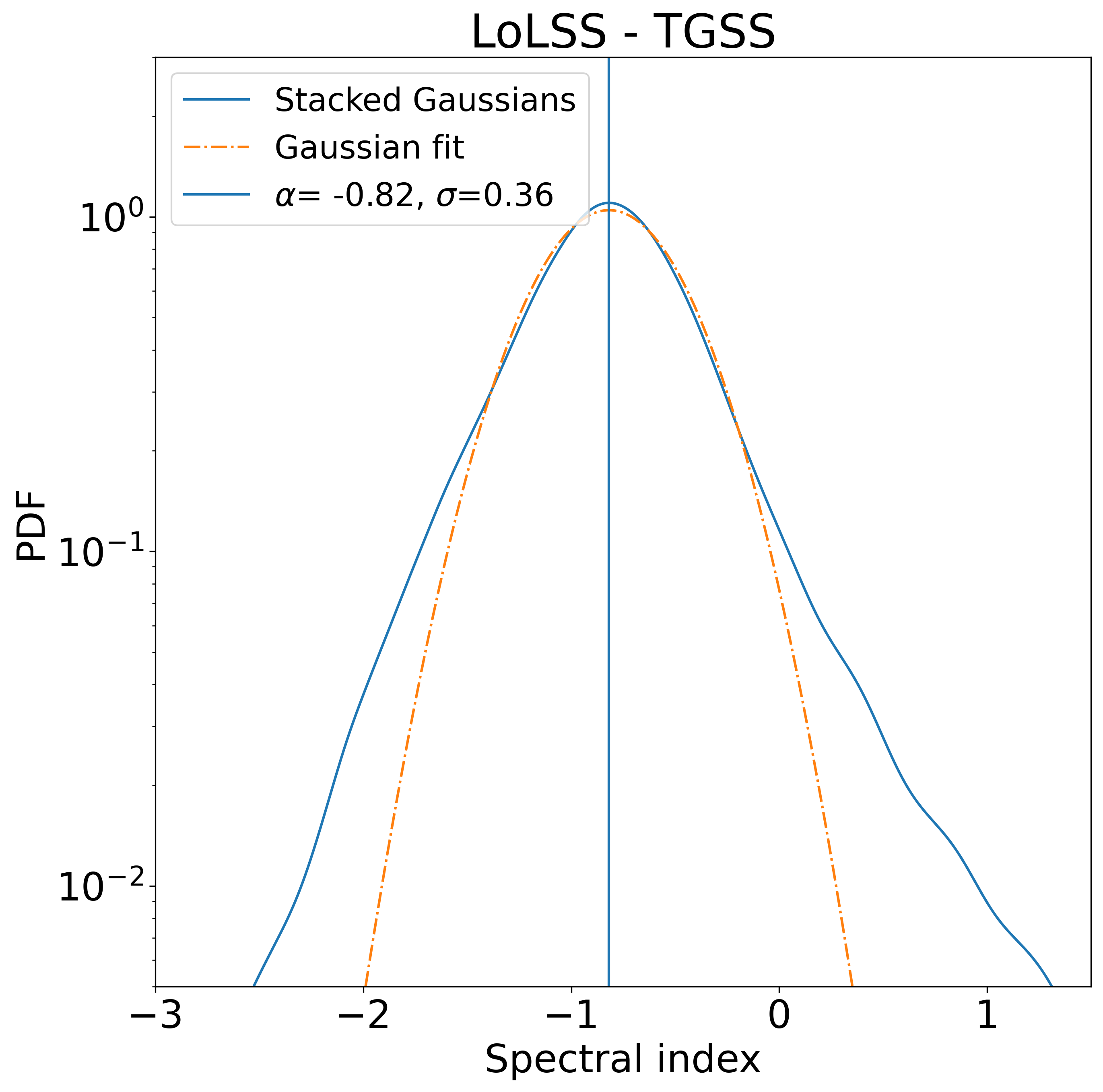}}
    \caption{Spectral index distribution for sources matched in LoLSS and TGSS. The orange line is a Gaussian fit with the parameters noted in the legend.}
    \label{fig:Ap_Alpha_LG}
\end{figure}
\begin{figure}[h!]
    \resizebox{\hsize}{!}{\includegraphics{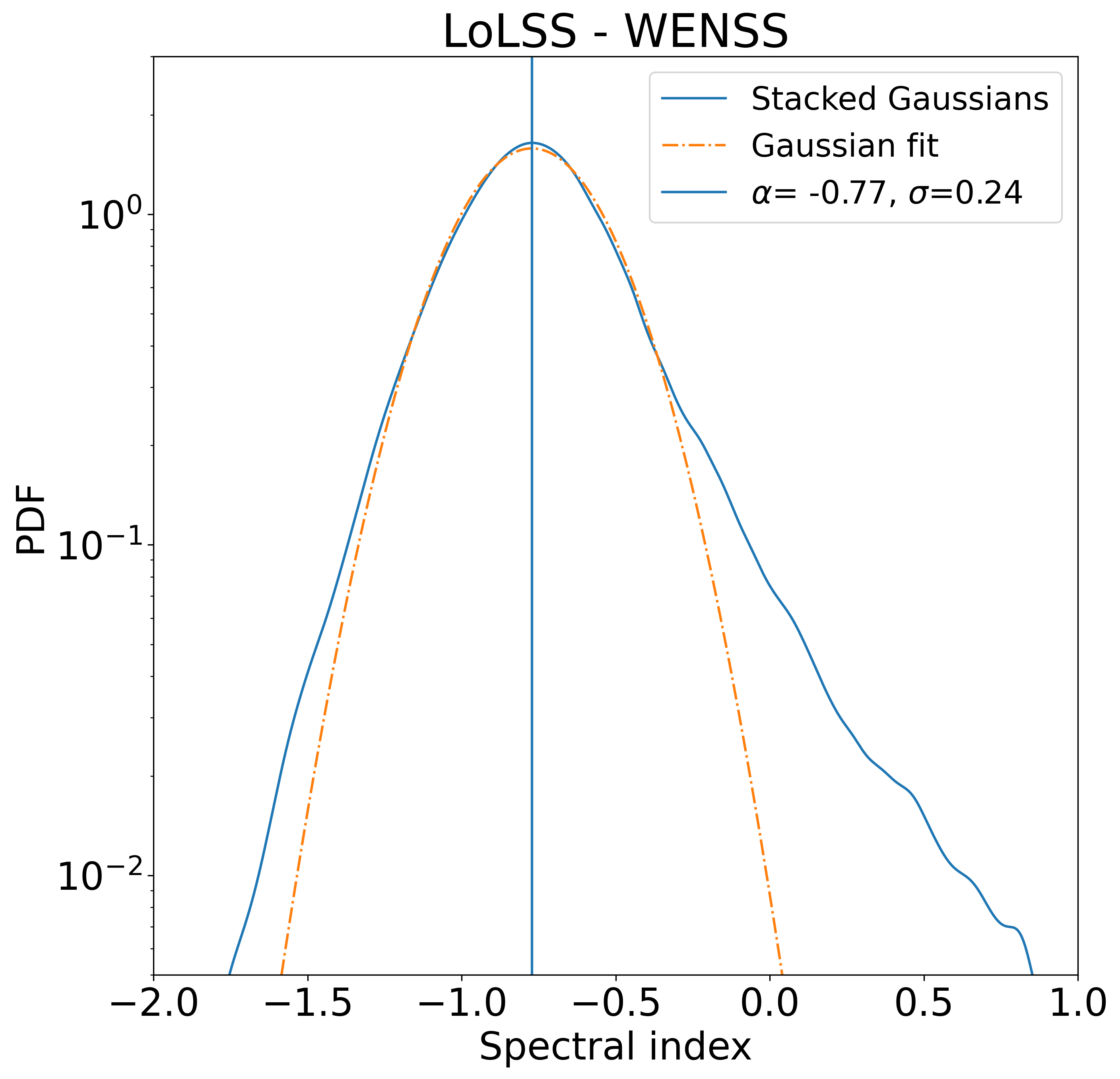}}
    \caption{Spectral index distribution for sources matched in LoLSS and WENSS. The orange line is a Gaussian fit with the parameters noted in the legend.}
    \label{fig:Ap_Alpha_LW}
\end{figure}
\begin{figure}[h!]
    \resizebox{\hsize}{!}{\includegraphics{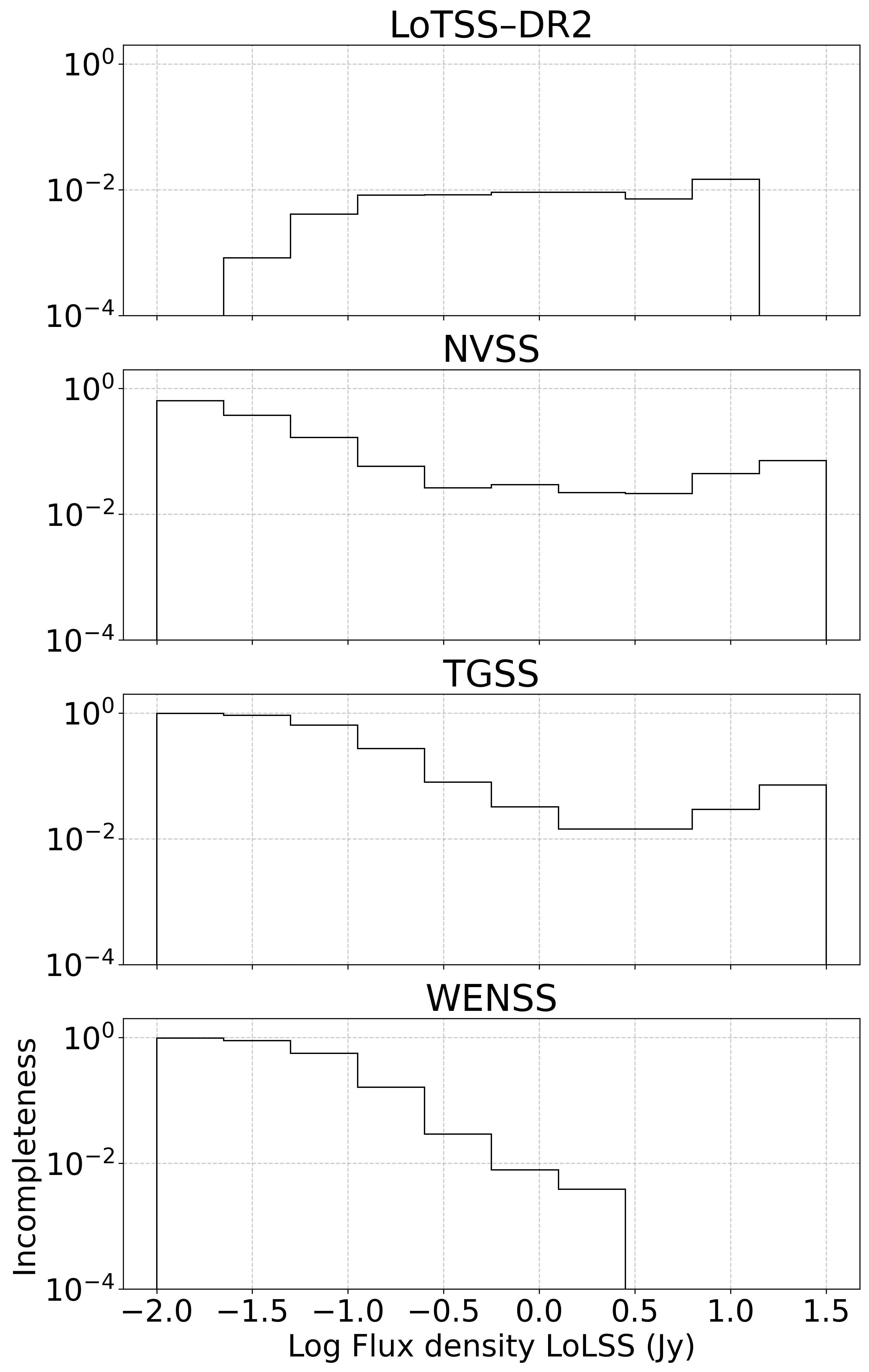}}
    \caption{Cross-matching incompleteness as a function of flux density for the different surveys (named above each plot) matched with LoLSS.}
    \label{fig:Ap_Incomplete}
\end{figure}
\begin{figure}[h!]
    \resizebox{\hsize}{!}{\includegraphics{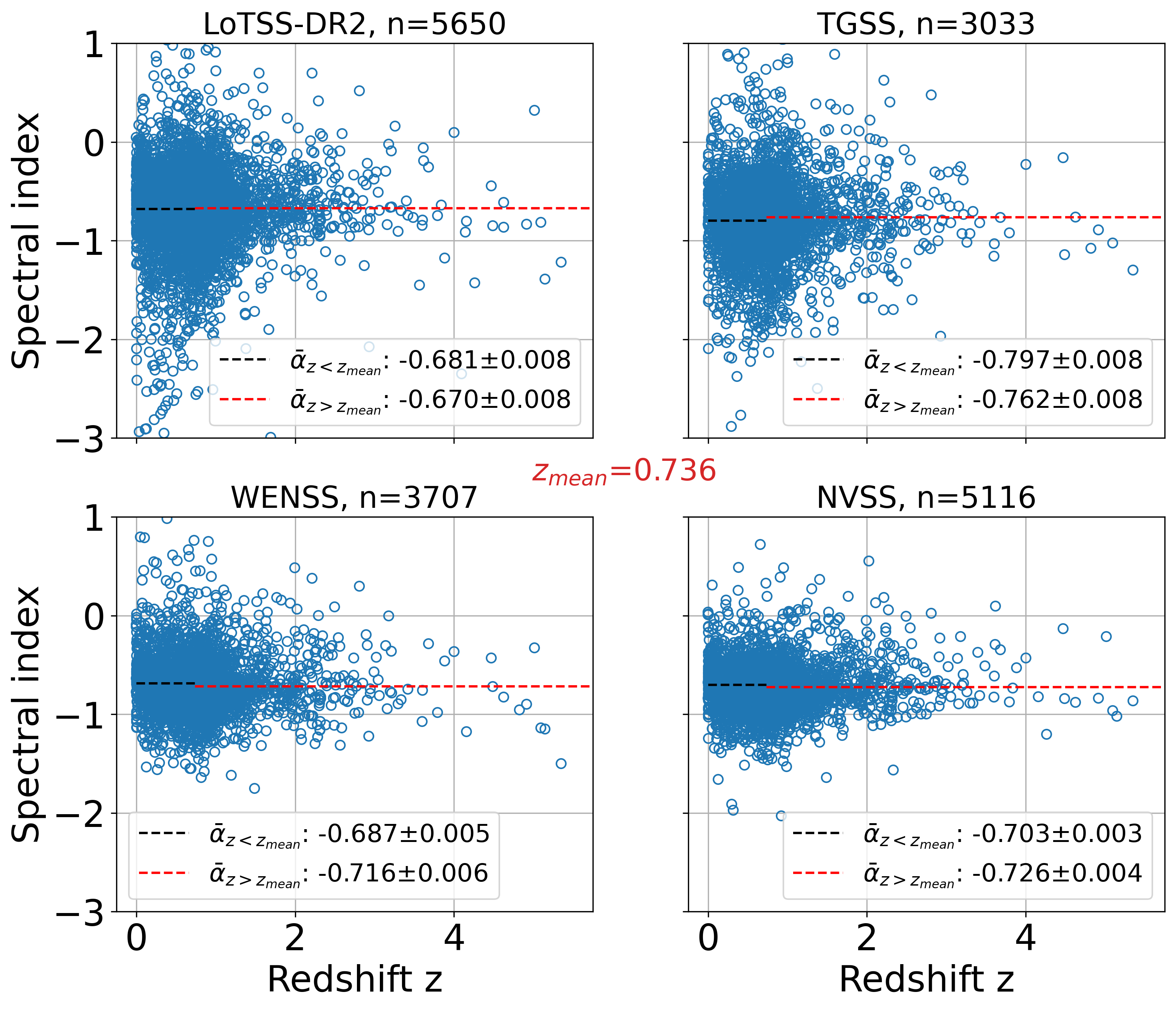}}
    \caption{Spectral index - Redshift relation. In each lower right corner is the mean spectral index in the two redshift bins above and below z$_{\mathrm{mean}}$, which itself is given in the centre. The black and red dotted lines show the two mean spectral indices.}
    \label{fig:redshift}
\end{figure}

\end{appendix}

\end{document}